\documentclass[acmtog, nonacm]{acmart}
\acmBooktitle{Research Report}

\usepackage{cleveref}
\usepackage{xcolor}

\usepackage{subcaption}
\usepackage{pgfplots}
\pgfplotsset{compat=1.17}
\usepackage{adjustbox}
\usepackage{multirow}
\usepackage{rotating}
\usepackage{wrapfig}
\usepackage{svg}
\usepackage{enumitem}
\usepackage[skins]{tcolorbox}
\usepackage{algorithm2e}
\RestyleAlgo{ruled}
\usepackage{dblfloatfix}
\usepackage{layouts}

\newcommand{\ignore}[1]{}
\crefformat{equation}{Eq.~(#2#1#3)}
\renewcommand{\eqref}{\cref}

\newcommand{\vecx}{\mathbf{x}}
\newcommand{\vecy}{\mathbf{y}}
\newcommand{\vecz}{\mathbf{z}}
\newcommand{\vecr}{\mathbf{r}}
\newcommand{\vecn}{\mathbf{n}}

\newcommand{\oH}{\operatorname{H}}
\newcommand{\oI}{\operatorname{I}}

\definecolor{tabred}{rgb}{0.839216,0.153941,0.156863}
\definecolor{tabblue}{rgb}{0.121569,0.466667,0.705882}

\title{A Practical Walk-on-Boundary Method for Boundary Value Problems}
\author{Ryusuke Sugimoto}
\affiliation{
    \institution{University of Waterloo}
    \country{Canada}
}
\email{rsugimot@uwaterloo.ca}
\author{Terry Chen}
\affiliation{
    \institution{University of Waterloo}
    \country{Canada}
}
\email{ty6chen@uwaterloo.ca}
\author{Yiti Jiang}
\affiliation{
    \institution{University of Waterloo}
    \country{Canada}
}
\email{yt2jiang@uwaterloo.ca}
\author{Christopher Batty}
\affiliation{
    \institution{University of Waterloo}
    \country{Canada}
}
\email{christopher.batty@uwaterloo.ca}
\author{Toshiya Hachisuka}
\affiliation{
    \institution{University of Waterloo}
    \country{Canada}
}
\email{toshiya.hachisuka@uwaterloo.ca}



\begin{document}
\begin{abstract}
We introduce the \emph{walk-on-boundary} (WoB) method for solving boundary value problems to computer graphics. 
WoB is a grid-free Monte Carlo solver for certain classes of second order partial differential equations. 
A similar Monte Carlo solver, the walk-on-spheres (WoS) method, has been recently popularized in computer graphics due to its advantages over traditional spatial discretization-based alternatives. 
We show that WoB's intrinsic properties yield further advantages beyond those of WoS.
Unlike WoS, WoB naturally supports various boundary conditions (Dirichlet, Neumann, Robin, and mixed) for both interior and exterior domains.
WoB builds upon boundary integral formulations, and it is mathematically more similar to light transport simulation in rendering than the random walk formulation of WoS. 
This similarity between WoB and rendering allows us to implement WoB on top of Monte Carlo ray tracing, and to incorporate advanced rendering techniques (e.g., bidirectional estimators with multiple importance sampling, the virtual point lights method, and Markov chain Monte Carlo) into WoB.
WoB does not suffer from the intrinsic bias of WoS near the boundary and can estimate solutions precisely on the boundary. 
Our numerical results highlight the advantages of WoB over WoS as an attractive alternative to solve boundary value problems based on Monte Carlo.
\end{abstract}

\maketitle

\begin{figure}[t]
\centering
\resizebox{\linewidth}{!}{
\includegraphics{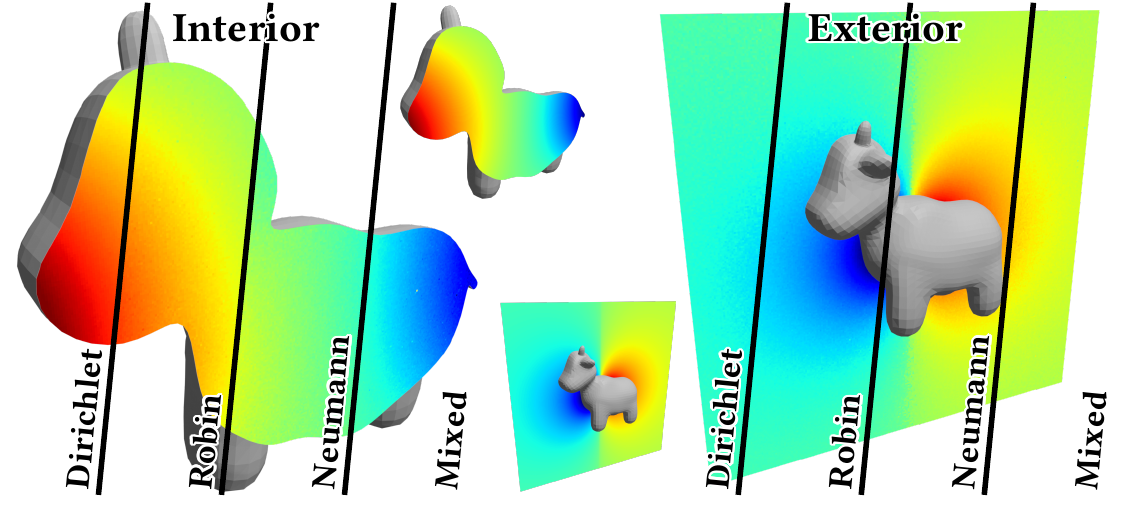}
}

\caption{The walk-on-boundary (WoB) method can handle various boundary value problems including Dirichlet, Robin, Neumann, and mixed for both interior and exterior problems under the same framework based on boundary integral equations. In this experiment, each problem is configured to have the same ground-truth solution (middle), and this figure shows that our WoB estimators all converge to that same solution (left and right).}
\label{fig:3d_teaser}
\end{figure}

\section{Introduction}\label{sec:introduction} 
Boundary value problems are relevant to diverse applications in computer graphics and beyond. 
Since the analytical solution is generally unavailable, one must resort to a numerical method to estimate the solution for practical problems. 
Such numerical methods are conventionally based on discretizing and then solving matrix equations. 
\citet{Sawhney:2020:MCG} recently introduced to computer graphics an alternative Monte Carlo (MC) solver called the \emph{walk-on-spheres} (WoS) method~\cite{Muller:1956:WoS}. 
They showed that WoS possesses various advantages over the conventional methods, including flexibility of geometric representation, robustness, parallelism, applicability to many geometric problems, and pointwise solution estimations.
The connection between WoS and MC ray tracing~\cite{Pharr:2018:PBRT} was also suggested by these authors and others \cite{Sawhney:2022:GFMC, Yilmazer2022,RiouxLavoie2022:Fluids}, which enables various techniques from MC ray tracing to be modified and adapted to WoS, such as in recent work by \citet{qi22bidirectional} on bidirectional WoS.

\begin{figure}[t]
\centering
\includegraphics{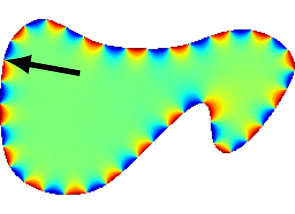}
\resizebox{0.6\linewidth}{!}{
\includegraphics{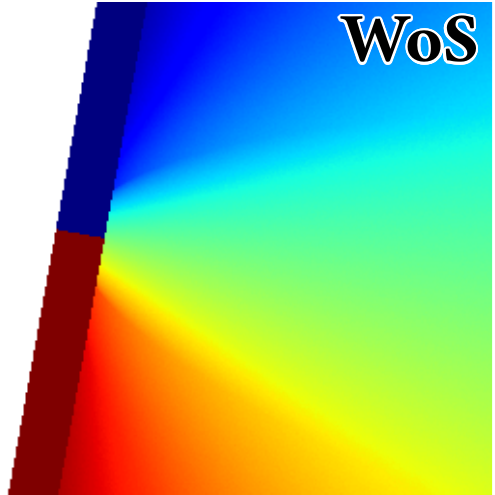}
\includegraphics{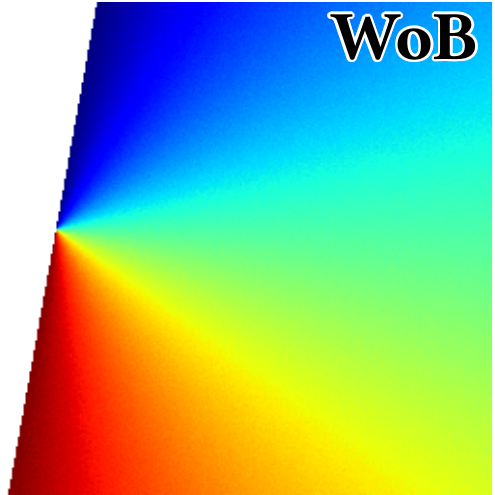}
}
\caption{Unlike WoS, WoB does not introduce any errors associated with the $\epsilon$-shell path termination. The indicated region (left) displays visible banding artifacts with WoS (middle), where WoB presents no such error (right).}%
\label{fig:nearboundry}
\end{figure}

\sloppy
We introduce another MC-based solver for boundary value problems: the \emph{walk-on-boundary} (WoB) method~\cite{sabelfeld1982vector,sabelfeld1991monte}.
Unlike WoS, which is based on random walks, WoB is based on \emph{potential theory}, which is the study of harmonic functions in mathematical physics. 
Potential theory allows us to convert the partial differential equations of boundary value problems into various forms of \emph{boundary integral equations} (BIEs). 
Just like the integral equation in rendering~\cite{Kajiya:1986:Rendering},
the resulting equations can be solved by tracing rays, rather than using random walks as in WoS.
WoB is arguably less widely known compared to WoS and %
we are the first to introduce WoB to computer graphics.
WoB and WoS share many advantages over spatial discretization-based methods, like finite element or finite difference methods, but WoB possesses further advantages that fundamentally differentiate it from WoS. We explore and demonstrate those advantages, specifically:
\begin{itemize}[leftmargin=*]
\item 
\textbf{Generality:} WoB can handle problems for which WoS is either inapplicable or inefficient.
Handling Neumann or Robin boundaries with WoS~\cite{Simonov2008, simonov2017walk} is known to be inefficient for non-convex domains. 
Only concurrent work by \citet{sawhney2023walk} offers an extension to WoS for efficiently handling Neumann boundaries.
Likewise, efficient handling of exterior problems in WoS requires Kelvin transformations into interior problems~\cite{Nabizadeh:2021:Kelvin}. 
By formulating the appropriate integral equation, our WoB solver handles these cases effortlessly, without changing the core algorithm.
Fig.~\ref{fig:3d_teaser} showcases this generality with results for Dirichlet, Neumann, Robin, and mixed boundary conditions for both interior and exterior domains.

\item \textbf{Accuracy:} WoB can accurately estimate the solution close to and even directly \emph{on} the boundary itself, where WoS becomes inefficient, inaccurate, or inapplicable. 
Fig.~\ref{fig:nearboundry} compares the accuracy of WoS and WoB near the boundary.
Solutions on the boundary are relevant for Neumann and Robin problems, since the boundary values are unknown, and they are the main quantity of interest for certain cases~\cite{Da2016, Sugimoto:2022:BEM}.

\item \textbf{Similarity to MC ray tracing:} WoB performs calculations by sampling points \emph{on} the boundary as in MC ray tracing; by contrast,
WoS is based on random walks \emph{inside} the boundary shape using closest point queries, which is not entirely equivalent to MC ray tracing.
Fig.~\ref{fig:overview} illustrates this difference. %
WoB is both mathematically and algorithmically more similar to MC rendering than WoS. 
This similarity allows WoB to leverage existing ray tracing frameworks and straightforwardly transform advanced rendering methods into solvers for boundary value problems.%
\end{itemize}

\begin{algorithm}[b]
    \DontPrintSemicolon
    \caption{Interior, convex-domain, and Dirichlet WoB}
    \label{alg:interior_dirichlet_convex}
    \SetKwInOut{Input}{Input}
        \Input{domain $\Omega$, evaluation point $\vecx_0\in \Omega$,\\path length $M$, sample count $N$}
    \vskip 0.5EM
    \SetKwFunction{FMain}{EstimateSolution}
    \SetKwProg{Fn}{Function}{:}{}
    \Fn{\FMain{$\Omega$, $\vecx_0$, $M$, $N$}}{
        $\hat{u}_{\text{sum}} \leftarrow 0$\\
        \For {$n\leftarrow 1$ to $N$}{
            $\hat{u}$ $\leftarrow$ \texttt{RecursiveEstimate} ($\Omega$, $\vecx_0$, $M$, $0$) \\
            $\hat{u}_{\text{sum}}$ $\leftarrow$$\hat{u}_{\text{sum}} + \hat{u}$\\
        }
        \Return $\hat{u}_{\text{sum}}/N$
    }
    \vskip 0.5EM
    \SetKwFunction{FMain}{RecursiveEstimate}
    \Fn{\FMain{$\Omega$, $\vecx_i$, $M$, \texttt{depth}}}{
        $\vecx_{i+1}$ $\leftarrow$ \texttt{RayIntersectionSampling}($\Omega$, $\vecx_i$)\\
        \uIf {$\texttt{depth} = M-1$}{\Return $\overline u_D(\vecx_{i+1})$}
        \Else {
        $\hat{u}_{i+1}$ $\leftarrow$ \texttt{RecursiveEstimate}($\Omega$, $\vecx_{i+1}$, $M$, $\text{\texttt{depth}}+1$)\\
        \Return{$2\overline u_D(\vecx_{i+1})- \hat{u}_{i+1}$}
        }
    }
    \vskip 0.5EM
    \SetKwFunction{FMain}{RayIntersectionSampling}
    \Fn{\FMain{$\Omega$, $\vecx$}}{
        $\mathbf{d}$ $\leftarrow$ \texttt{UniformRayDirectionSampling}($\Omega$, $\vecx$)\\
        \Return \texttt{GetIntersectionPoint}($\Omega$, \texttt{Ray}($\vecx$, $\mathbf{d}$))
    }
\end{algorithm}

\section{Overview}
\label{sec:overview}
Let us examine instances of WoS and WoB estimators to highlight the differences. 
Consider a boundary value problem in which a function $\overline u_D(\vecx)$ is specified along the boundary of a convex domain $D$, for simplicity. We will relax this assumption of convexity in Section~\ref{sec:wob}.
We seek a function $u(\vecx)$ that conforms to $\overline u_D(\vecx)$ on the boundary while satisfying Laplace's equation, $\Delta u(\vecx) = 0$, inside the domain. \looseness=-1

The WoS estimator in this case is defined as the mean of sample contributions $\hat u(\vecx_0)$ defined recursively as
\begin{equation}
\hat u(\vecx_i) = \begin{cases} \overline u_D(\vecx_i) &\quad \vecx_i \in \partial D_{\epsilon}, \\
\hat u(\vecx_{i+1}) &\quad \mathrm{otherwise.} \end{cases}
\end{equation}
Starting from an evaluation point $\vecx_0$ within the domain, 
each subsequent $\vecx_{i+1}$ is generated by sampling a point on the largest ball centered at $\vecx_i$ that fits within the domain.
The set $\partial D_{\epsilon}$ is a shell-like region that lies within a small distance of the boundary. 
WoS thus returns $\overline{u}_D(\vecx_i)$  when $\vecx_i$ is only \emph{approximately} on the boundary; indeed, none of the generated points are exactly on the boundary.

\begin{figure}[t]
\includegraphics{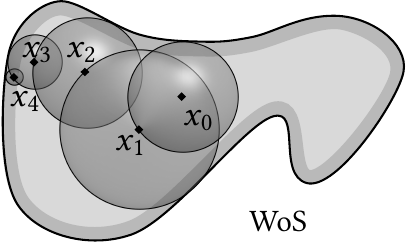}
\includegraphics{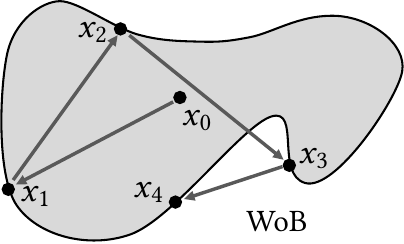}
\caption{While WoS takes random walks on spheres, WoB takes random walks on the boundary to compute a sample contribution.}
\label{fig:overview}
\end{figure}

The WoB estimator with path length $M$ we introduce is defined~as
\begin{equation}
\hat u(\vecx_i) = \begin{cases} \overline{u}_D(\vecx_{i+1}) &\quad i=M-1, \\
 2 \overline{u}_D(\vecx_{i+1}) - \hat u(\vecx_{i+1}) &\quad \mathrm{otherwise.} \end{cases}
\end{equation}
Like WoS, WoB forms a sequence of $\vecx_i$ starting from the evaluation point $\vecx_0$. 
Each $\vecx_{i+1}$, however, is generated by tracing a random ray from $\vecx_k$ and finding the intersection with the boundary. 
Unlike WoS, all the points (after $\vecx_0$) are \emph{exactly} on the boundary, and thus $\overline u_D(\vecx_{i+1})$ is well-defined, without approximation via the $\epsilon$-shell $\partial D_{\epsilon}$. 
As in MC ray tracing, the recursion terminates after some number of steps $M$ and does not depend at all on the $\epsilon$-shell $\partial D_{\epsilon}$. 
Algorithm~\ref{alg:interior_dirichlet_convex} summarizes this method and %
Fig.~\ref{fig:overview} illustrates how sequences of $\vecx_i$ differ between WoS and WoB in a non-convex domain.

Just like MC ray tracing, in both WoS and WoB, one would run the estimator multiple times and take its average as a final estimation of $u(\vecx_0)$.
However, because WoB is also based on sampling and tracing rays recursively, the analogy with MC ray tracing is much stronger, in sharp contrast to the random walk on spheres process used by WoS.
One of our contributions is to demonstrate WoB's strong mathematical and algorithmic similarity to MC ray tracing.

The ray tracing approach to diffusion curves~\cite{Bowers2011diffusion} also solves a boundary integral and estimates the solution by sampling points on the boundary using ray intersections, similarly to WoB. 
While this approach provides a \emph{visually} faithful approximation 
for diffusion curves~\cite{Orzan2008}, it does not actually solve Laplace's equation. 
By contrast, the WoB formulation does solve the specified PDE (i.e., Laplace's equation) and can handle more general problems, such as Neumann problems.

\section{Boundary Integral Formulations} \label{sec:background}
WoB builds upon boundary integral equation formulations.
Such formulations are applicable to many second-order linear elliptic PDEs~\cite{Clements2004} and some other PDEs~\cite{Liu2012}, but we focus on boundary value problems based on Poisson's and Laplace's equations.
The book by \citet{SabelfeldSimonov1994} and notes by \citet{Clemens:BEM2013} discuss the details of BIEs summarized here.%

\subsection{Poisson's Equation and Boundary Conditions} 
Poisson's equation is commonly given as
\begin{equation}\label{eq:Poisson}
    \Delta v(\vecx) = \overline b(\vecx) \quad\text{for}\quad \vecx\in \Omega
\end{equation}
where $\Omega$ is a closed domain in $\mathbb{R}^2$ or $\mathbb{R}^3$, $v(\vecx)$ is the unknown field, and $\overline b(\vecx)$ is a known source function defined inside the domain. 
Poisson's equation with $\overline b(\vecx) = 0$ is also called Laplace's equation.
We will later consider exterior problems where the domain is $\mathbb{R}^2\backslash\Omega$ or $\mathbb{R}^3\backslash\Omega$. 
Let us denote the boundary of $\Omega$ by $\Gamma = \partial \Omega$; note that $\Gamma$ is not included in $\Omega$. 
The \emph{boundary value problem} in this paper concerns solving for $v(\vecx)$ in \eqref{eq:Poisson} with boundary conditions
\begin{align}\label{eq:PoissonBC}
    v(\vecx) &= \overline u_D(\vecx) &\text{ }&\vecx \in \Gamma_D \subseteq \Gamma,\nonumber \\
    \frac{\partial v}{\partial \vecn}(\vecx) &= \overline q_N(\vecx) &\text{ }&\vecx \in \Gamma_N \subseteq \Gamma,\text{  }\\
    \frac{\partial v}{\partial \vecn}(\vecx) + \overline\alpha(\vecx) v(\vecx) &= \overline g_R(\vecx) &\text{ }& \vecx \in \Gamma_R \subseteq \Gamma,\nonumber 
\end{align}
where the normal derivative $\frac{\partial v}{\partial \vecn}(\vecx) =  \vecn(\vecx) \cdot \nabla v(\vecx)$, the weight $\overline\alpha(\vecx)\neq0$, and a bar indicates that the function is given. 
Any point $\vecx\in\Gamma$ belongs to strictly one of $\Gamma_D, \Gamma_N$, or $\Gamma_R$.
When  $\Gamma_D = \Gamma$, $\Gamma_N = \Gamma$, or $\Gamma_R = \Gamma$, we call the problem a Dirichlet, Neumann, or Robin problem, respectively. Otherwise, we call it a mixed boundary problem.

One can always convert Poisson's equation for $v(\vecx)$ into Laplace's equation for $u(\vecx)$ using a relation $u(\vecx) = v(\vecx) + V_0(\vecx)$ as follows. 
Let $u$ be a function satisfying Laplace's equation
\begin{equation}\label{eq:Laplace}
\begin{aligned}
  \Delta u(\vecx) &= 0 &\text{ }&\vecx\in \Omega\\
    u(\vecx) &= \overline u_D(\vecx) + V_0(\vecx) &\text{ }& \vecx \in \Gamma_D,\\
    \frac{\partial u}{\partial \vecn}(\vecx) &= \overline q_N(\vecx) + \frac{\partial V_0}{\partial \vecn} (\vecx) &\text{ }& \vecx \in \Gamma_N,  \\
    \frac{\partial u}{\partial \vecn}(\vecx) + \overline\alpha(\vecx) u(\vecx) &= \overline g_R(\vecx) +  \frac{\partial V_0}{\partial \vecn} (\vecx) + \overline\alpha(\vecx)V_0(\vecx) &\text{ }& \vecx \in \Gamma_R,\\
\end{aligned}
\end{equation}
where
$
    V_0(\vecx) = \int_\Omega G(\vecx, \vecy) \overline b(\vecy) \,dV_\vecy
$
and $G(\vecx, \vecy)$ is the fundamental solution (i.e., the solution to Poisson's equation in an infinite domain with a negative Dirac delta source at each point). 
We list the fundamental solution and their derivatives used in this paper in Appendix~\ref{app:fund_sol} for readers' convenience. 
One can confirm that $v(\vecx) = u(\vecx) - V_0(\vecx)$ satisfies Poisson's equation by noting that $\Delta V_0(\vecx) = -\overline b(\vecx)$ for $\vecx \in \Omega$. 
We focus on solutions to Laplace's equation since it is easy to handle $V_0(\vecx)$, as we discuss later.

\subsection{Direct Boundary Integral Equations}\label{sec:directBIE}
The basic idea of boundary integral formulations is to define the solution based on integrals \emph{only of boundary values}.
They can actually take several different forms.  
We first explain one such formulation called a \emph{direct} BIE formulation. 
As the name suggests, direct BIEs describe the relationship between the solution values in the interior or on the boundary \emph{directly}, and are derived
based on Green's third identity.
One common such direct BIE is given as %
\begin{equation}\label{eq:cbie}
\begin{split}
c(\vecx)u(\vecx) = &-  \int_\Gamma \frac{\partial G}{\partial \vecn_\vecy}(\vecx, \vecy) u(\vecy) \,dA_\vecy + \int_\Gamma G(\mathbf{x}, \mathbf{y}) \frac{\partial u}{\partial \vecn} (\vecy) \,dA_\vecy\\
\end{split}
\end{equation}
for $\vecx \in \Omega \cup \Gamma$, where $c(\vecx)$ is an integral free term that evaluates to $1$ if $\vecx \in \Omega$, and to $\tfrac{1}{2}$ if $\vecx \in \Gamma$ when the boundary is smooth in the sense of Lyapunov. 
We consider only smooth surfaces in the following, but the extension to non-smooth surfaces is straightforward.
Note that polygonal boundaries do not violate the smoothness assumption, unless $\vecx$ is evaluated \emph{exactly} on a vertex of the polygon. 

\eqref{eq:cbie} says that the solution $u(\vecx)$ satisfies this integral equation involving only boundary integrals.
The boundary conditions alone do not provide $u(\vecy)$ and $\frac{\partial u}{\partial \vecn} (\vecy)$ everywhere along $\Gamma$, so one must solve for such unknown boundary values first to evaluate the solution $u(\vecx)$ inside of the domain. 
Concurrent work by \citet{miller2023boundary} uses this equation to cache WoS (and Walk-on-Stars~\cite{sawhney2023walk}) estimates along the boundary to accelerate its computation; our WoB readily supports an analogous caching scheme.%

One can also derive a BIE for the directional derivative of the solution by taking the directional derivative of \eqref{eq:cbie}: 
\begin{equation}\label{eq:hbie2_regularized}
\frac{\partial u}{\partial \vecx_k}(\vecx) = - \int_\Gamma  \frac{\partial^2 G}{\partial \vecx_k\partial \vecn_\vecy}(\mathbf{x}, \mathbf{y}) u(\vecy)\,dA_\vecy\\
+ \int_\Gamma \frac{\partial G}{\partial \vecx_k}(\mathbf{x}, \mathbf{y})\frac{\partial u}{\partial \vecn}(\vecy) \,dA_\vecy,
\end{equation}
for $\vecx \in \Omega$. We denote the first order derivative with respect to the $k$-th direction by $\partial u/\partial \vecx_k$. This equation is valid only in the interior of the domain, so we omitted the integral free term $c(\vecx)=1$. %

\subsection{Indirect Boundary Integral Equations}
In contrast to direct BIEs, an \emph{indirect} BIE describes the relationship between an unknown \emph{source density function} on the boundary and the known boundary values, and expresess the solution only \emph{indirectly} based on the source density function. %
There are two types of indirect formulations, derived from potential theory.

\paragraph{Single layer potential} 
The solution $u(\vecx)$ to Laplace's equation can be expressed in the form of a \emph{single layer potential} given by
\begin{equation}\label{eq:indirect_single_u}
    u(\vecx) = \int_\Gamma G(\vecx, \vecy) \mu(\vecy) \,dA_\vecy \quad\text{for}\quad \vecx\in \Omega \cup \Gamma,\\
\end{equation}
where the unknown source density function $\mu$ corresponds to the jump of the normal derivative of $u$ across the boundary.
In short, \eqref{eq:indirect_single_u} expresses the solution inside the domain (and on the boundary) in terms of monopole sources, distributed over the boundary, which decay according to the fundamental solution $G(\vecx, \vecy)$. %
Taking the directional derivative of \eqref{eq:indirect_single_u} and taking the limit to the boundary gives an integral equation for $\mu(\vecx)$: %
\begin{equation}\label{eq:indirect_single_q}
\frac{\partial u}{\partial \vecn}(\vecx) =  c(\vecx)\mu(\vecx)  + \int_\Gamma \frac{\partial G}{\partial \vecn_\vecx}(\vecx, \vecy) \mu(\vecy) \,dA_\vecy \quad\text{for}\quad \vecx\in \Gamma.
\end{equation}
Similarly, the directional derivative at any interior point $\vecx$ is
\begin{equation}\label{eq:indirect_single_dudx}
    \frac{\partial u}{\partial \vecx_k}(\vecx) = \int_\Gamma \frac{\partial G}{\partial \vecx_k}(\vecx, \vecy) \mu(\vecy) \,dA_\vecy \quad\text{for}\quad \vecx\in \Omega.\\
\end{equation}
This equation is invalid exactly on the boundary because of jump discontinuities across $\Gamma$.

\paragraph{Double layer potential} An alternative is a \emph{double layer potential} which uses dipole source on the boundary:
\begin{equation}\label{eq:indirect_double_domain}
    u(\vecx) = -\int_\Gamma \frac{\partial G}{\partial \vecn_\vecy}(\vecx, \vecy) \nu(\vecy) \,dA_\vecy \quad\text{for}\quad \vecx\in \Omega 
\end{equation}
where $\nu$ is an unknown source density function, corresponding to the jump of the solution across the boundary.
In the limit as $\vecx \rightarrow \Gamma$, one finds an integral equation for $\nu(\vecx)$: 
\begin{equation}\label{eq:indirect_double_boundary}
u(\vecx) = [1-c(\vecx)]\nu(\vecx)  -\int_\Gamma \frac{\partial G}{\partial \vecn_\vecy}(\vecx, \vecy) \nu(\vecy) \,dA_\vecy \quad\text{for}\quad \vecx\in \Gamma.
\end{equation}
The normal derivative can be computed with
\begin{equation}\label{eq:indirect_double_dudn}
    \frac{\partial u}{\partial \vecn_\vecx}(\vecx) = -\frac{\partial}{\partial \vecn_\vecx}\int_\Gamma \frac{\partial G}{ \partial \vecn_\vecy}(\vecx, \vecy) \nu(\vecy) \,dA_\vecy \quad\text{for}\quad \vecx\in \Gamma,
\end{equation}
and the directional derivative can be computed with
\begin{equation}\label{eq:indirect_double_dudx}
    \frac{\partial u}{\partial \vecx_k}(\vecx) = -\int_\Gamma \frac{\partial^2 G}{\partial \vecx_k \partial \vecn_\vecy}(\vecx, \vecy) \nu(\vecy) \,dA_\vecy \quad\text{for}\quad \vecx\in \Omega.
\end{equation}

Having outlined BIE formulations for Laplace's equation, we proceed to solve them via WoB. 
To develop practical numerical WoB methods and also generalize WoB to mixed boundary problems, we will adopt different formulations for different problems.

\section{The Walk-on-Boundary Method} \label{sec:wob}
WoB is a stochastic estimator for certain classes of second-order PDEs based on BIEs.
It uses a sequence of stochastically chosen sample points on the boundary, hence the name "walk-on-boundary". 
\citet{sabelfeld1982vector} first proposed WoB for the Lam\'e equation for linear elasticity. Subsequent books by \citet{sabelfeld1991monte} and \citet{SabelfeldSimonov1994} generalized the method and established theoretical foundations for Dirichlet, Neumann, and Robin problems for Poisson's equation using the indirect BIE formulation, along with extensions to a few other equations.
\citet{KaraivanovaMascagniSimonov+2004+311+319} considered the applicability of quasi-Monte Carlo methods and \citet{SABELFELD2012} studied how WoB can be combined with the method of fundamental solutions to improve its efficiency. %

None of this prior work has considered the application of WoB ideas to direct BIE formulations or mixed boundary problems; these extensions are our contributions.
We also have identified that using double layer potential indirect BIEs for Dirichlet boundary problems, single layer potential indirect BIEs or direct BIEs for Neumann boundary problems, and single layer potential indirect BIEs for Robin and mixed boundary problems will result in practical solvers.
We summarize the equations for formulations for all of our estimators discussed in this paper in Table~\ref{table:equations}.

\newcommand{\domaincoeff}{\phi}
\renewcommand{\arraystretch}{1.5}
\begin{table*}[t]
\centering
\caption{List of equations for WoB estimators. The highlighted equations are the second kind Fredholm equations (or the modified first-kind equation for mixed boundary problems at Dirichlet boundaries) we use to get the unknown direct or indirect quantities on the boundary. It can be combined with other equations to find the unknown quantities of interest in the interior (or exterior) or on the boundary. It is assumed that the left hand side unknowns are functions of $\vecx$, and the integrals are taken over boundary points $\vecy$. The explicit dependencies on variables are omitted for brevity when it is not confusing. For interior problems, $\domaincoeff=1$, and for exterior problems, $\domaincoeff=-1$.}
\begin{tabular}{|l|l|l|l|l|}
\hline
Problem&Formulation&Quantity to Estimate&Evaluation Point $\vecx$&Equation\\
\hline\hline
\multirow{5}{*}{Dirichlet}&
\multirow{5}{*}{\shortstack[l]{indirect BIE\\double layer\\potential}}
&\multirow{2}{*}{solution}&interior/exterior&$u = -\int \frac{\partial G}{\partial \vecn_\vecy}  \nu \,dA$\\
\cline{4-5}
&&&boundary & $u=\overline{u}_D$ (given)\\
\cline{3-5}
&&normal derivative&boundary& Sect. 3 in the book by \citet{SabelfeldSimonov1994}.\\%$\frac{\partial u}{\partial \vecn_\vecx} = -\int \frac{\partial^2 G}{\partial\vecn_\vecx\partial\vecn_\vecy}  \nu \,dA$ (\ryusuke{Not sure if $\phi$ is required here.}See comments in \S\ref{sec:misc_notes}.)\\
\cline{3-5}
&&gradient&interior/exterior& $\frac{\partial u}{\partial \vecx_k} = -\int \frac{\partial^2 G}{\partial \vecx_k\partial \vecn_\vecy}  \nu \,dA$ \\
\cline{3-5}
&&\textbf{source density}&\textbf{boundary}& $\nu = 2\domaincoeff\int \frac{\partial G}{\partial \vecn_\vecy} \nu \,dA  + 2\domaincoeff\overline{u}_D$\\
\hline
\multirow{4}{*}{Neumann}&
\multirow{4}{*}{direct BIE}
&\multirow{2}{*}{\textbf{solution}}&interior/exterior&$u =   - \domaincoeff\int \frac{\partial G}{\partial \vecn_\vecy}u \,dA +  \domaincoeff\int  G \overline{q}_N  \,dA$\\
\cline{4-5}
&&&\textbf{boundary} & $u =   -  2\domaincoeff\int \frac{\partial G}{\partial \vecn_\vecy}u \,dA +   2\domaincoeff\int G \overline{q}_N  \,dA$\\
\cline{3-5}
&&normal derivative&boundary & $\frac{\partial u}{\partial \vecn} = \overline{q}_N$ (given)\\
\cline{3-5}
&&gradient&interior/exterior&$\frac{\partial u}{\partial \vecx_\mathsf{k}} =  - \domaincoeff\int  \frac{\partial^2 G}{\partial \vecx_\mathsf{k}\partial \vecn_\vecy} u
\,dA + \domaincoeff\int \frac{\partial G}{\partial \vecx_\mathsf{k}}\overline{q}_N\,dA$\\
\hline
\multirow{6}{*}{\shortstack[l]{Mixed or\\degenerate\\problem}}&\multirow{6}{*}{\shortstack[l]{indirect BIE\\single layer\\potential}}
&solution&interior/exterior/boundary& $u = \int G \mu \,dA$\quad(or given)\\
\cline{3-5}
&&normal derivative& boundary &$\frac{\partial u}{\partial \vecn} = \frac{1}{2}\domaincoeff\mu + \int \frac{\partial G}{\partial \vecn_\vecx} \mu \,dA$\quad(or given)\\
\cline{3-5}
&&gradient&interior/exterior& $\frac{\partial u}{\partial \vecx_\mathsf{k}} = \int \frac{\partial G}{\partial \vecx_\mathsf{k}} \mu \,dA$\\
\cline{3-5}
&&\multirow{3}{*}{\textbf{source density}}&\textbf{Dirichlet boundary}& $\mu = \mu - k\int G\mu \,dA + k\overline{u}_D$\\
\cline{4-5}
&&&\textbf{Neumann boundary}& $\mu =  -  2\domaincoeff\int\frac{\partial G}{\partial \vecn_\vecx} \mu \,dA + 2\domaincoeff\overline{q}_N $ \\
\cline{4-5}
&&&\textbf{Robin boundary}& $\mu =  -  2\domaincoeff\int\left(\frac{\partial G}{\partial \vecn_\vecx}+\overline\alpha(\vecx)G\right) \mu \,dA + 2\domaincoeff\overline{g}_R $ \\
\hline
\end{tabular}
\label{table:equations}
\end{table*}

\subsection{Dirichlet Problems with Double Layer BIE}\label{sec:DirichletDoubleLayer}
For Dirichlet problems, reordering terms in  \eqref{eq:indirect_double_boundary} and substituting in the boundary condition $u=\overline u_D$ gives
\begin{equation}\label{eq:indirect_double_boundary_reordered}
\nu(\vecx) = \int_\Gamma 2\frac{\partial G}{\partial \vecn_\vecy}(\vecx, \vecy) \nu(\vecy) \,dA_\vecy  + 2\overline{u}_D(\vecx)\quad\text{for}\quad \vecx\in \Gamma,
\end{equation}
assuming that $\vecx$ lies on a smooth boundary ($c = 1/2$).
Since $\overline{u}_D(\vecx)$ is a known quantity, \eqref{eq:indirect_double_boundary_reordered} is a Fredholm equation of the second kind for $\nu$, as the rendering equation is also commonly understood to be. Appendix~\ref{sec:classes} elaborates on this point.
We can thus apply a recursive estimate for $\nu(\vecx)$ similarly to what is done for the rendering equation in light transport simulation~\cite{Pharr:2018:PBRT}. 

\subsubsection{MC estimation} 
Let us consider estimating $\nu(\vecx_1)$ where $\vecx_1 \in \Gamma$ based on MC integration.
We can estimate the integral in \eqref{eq:indirect_double_boundary_reordered} via MC integration by first sampling a point $\vecx_2\in\Gamma$ with a probability density function (PDF) $p(\vecx_2|\vecx_1)$ (e.g., tracing a random ray from $\vecx_1$ to $\vecx_2$).
A sample $\hat{\nu}(\vecx_1)$ to estimate ${\nu}(\vecx_1)$ can be written as
\begin{equation}
\hat{\nu}(\vecx_1) \coloneqq \frac{2\frac{\partial G}{\partial \vecn_\vecy}(\vecx_1, \vecx_2)}{p(\vecx_2|\vecx_1)} \nu(\vecx_2) + 2\overline{u}_D(\vecx_1)\quad\text{for}\quad \vecx_1\in \Gamma.
\end{equation}
Because $\nu(\vecx_2)$ is unknown, we again use an MC estimate $\hat\nu(\vecx_2)$ in the equation above. Thus a recursive definition for the $i$-{th} step is:
\begin{equation}\label{eq:indirect_double_boundary_reordered2}
\hat{\nu}(\vecx_i) \coloneqq \frac{2\frac{\partial G}{\partial \vecn_\vecy}(\vecx_i, \vecx_{i+1})}{p(\vecx_{i+1}|\vecx_i)} \hat{\nu}(\vecx_{i+1}) + 2\overline{u}_D(\vecx_i)\quad\text{for}\quad \vecx_i\in \Gamma.
\end{equation}
Just like MC ray tracing, we perform the recursive estimate of $\nu(\vecx)$ up to a certain recursion depth $M$, forming a path of vertices on the boundary with length $M$. 
We can use $\hat{\nu}(\vecx_1)$ to construct an MC estimate for the solution $u$ at an interior point $\vecx_0$. 
Applying another MC integration with a PDF $p(\vecx_1|\vecx_0)$ and $\hat{\nu}(\vecx_1)$ to \eqref{eq:indirect_double_domain} gives
\begin{equation}\label{eq:mc_dirichlet_interior_sol}
    \hat{u}(\vecx_0) \coloneqq -\frac{\frac{\partial G}{\partial \vecn_\vecy}(\vecx_0, \vecx_1) }{p(\vecx_1|\vecx_0)} \hat{\nu}(\vecx_1)\quad\text{for}\quad\vecx_0\in\Omega.
\end{equation}
Therefore, the MC estimate for $u(\vecx_0)$ in the domain interior is
$
    u(\vecx_0) \approx \frac{1}{N}\sum_{n=1}^N\hat{u}(\vecx_0).
$
One can think of the PDF $p(\vecx_{i+1}|\vecx_i)$ as the PDF of sampling a ray from $\vecx_i$ toward $\vecx_{i+1}$, the term $2\frac{\partial G}{\partial \vecn_\vecy}(\vecx_i, \vecx_{i+1})$ as the geometry term times BRDF term (i.e., the integrand) of the rendering equation, and the term $2\overline{u}_D(\vecx_i)$ as the emission term in the rendering equation. 
Implementation of WoB on top of ray tracing systems is thus straightforward.

\subsubsection{Path truncation}
One difference between the rendering equation and the above BIE is that, for the rendering equation, as the path length increases, the contribution coming from each recursion becomes smaller and smaller due to the nature of light transport.
However, the integral kernel $2\frac{\partial G}{\partial \vecn_\vecy}$ above will not "attenuate" its contribution per recursion, but will rather maintain it.
It thus appears that it never converges, just like having reflectance equal to one everywhere does not converge in light transport. 
This intuition contradicts with the fact that solutions usually uniquely exist for Dirichlet problems and it is incorrect.

A solution to this issue is surprisingly simple. 
We just need to multiply the contribution $2\overline{u}_D(\vecx_i)$ coming from the last recursion step by a factor of $1/2$ before it terminates:
$\hat{\nu}(\vecx_{M}) \coloneqq \overline{u}_D(\vecx_M)$. 
While this strategy deceivingly looks the same as just truncating a path while reducing the contribution from the last "bounce", its derivation is more involved than that. 
Below, we briefly summarize the rough reasoning behind this strategy.

Let us first define an integral operator $\oH$ when applied to a function $f$ defined over the boundary as
\begin{equation}
    (\oH f)(\vecx) = \int_\Gamma 2\frac{\partial G}{\partial \vecn_\vecy}(\vecx, \vecy) f(\vecy) \,dA_\vecy.
\end{equation}
Then, \eqref{eq:indirect_double_boundary_reordered} can be rewritten as 
\begin{equation}
    \nu = \oH\nu + 2\overline{u}_D,
\end{equation}
where we dropped the variable dependence for brevity. Using the identity operator $\oI$, we can write the expression above as
\begin{equation}
    (\oI - \oH)\nu =  2\overline{u}_D.
\end{equation}
One can use Neumann series expansion to solve for $\nu$ as
\begin{equation}
   \nu =   (\oI - \oH)^{-1}\;2\overline{u}_D
        =  (\oI + \oH + \oH^2 + \cdots)\;2\overline{u}_D, 
\end{equation}
where the operator $\oH^i$ for any positive integer $i$ is defined by
\begin{equation}
 \oH^i  f = \oH^{i-1}(\oH  f).
\end{equation}
The same approach is used for building a recursive MC estimator for the rendering equation where the operator is defined by BRDF and the geometry term instead~\cite{Pharr:2018:PBRT}.

As noted earlier, the key difference from the rendering equation is that $(\oH^i) 2\overline{u}_D(\vecx)$ does not approach zero as $i$ increases. 
Simply truncating this series at $M$ thus introduces non-negligible truncation error. 
We instead transform the series as
\begin{equation}
\begin{split}
\left(\frac12 (\oI + \cdots + \oH^{i-1}  + \cdots) + \frac12 (\oI + \cdots + \oH^{i}  + \cdots ) \right)2\overline{u}_D(\vecx) \\
= \frac12(\oI + (\oI + \oH) + \cdots + (\oH^{i-1} + \oH^{i}) + \cdots ) 2\overline{u}_D(\vecx).
\end{split}
\end{equation}
Just like multiple bounces in light transport, the average of the integrals $(1/2)(\oH^{i} + \oH^{i+1}) 2\overline{u}_D(\vecx)$ now converges to zero as $i \rightarrow \infty$ due to the alternating sign in the series because of the negative factor included in the operator $\oH$, so it is safe to truncate this modified series at $i=M$:
\begin{equation}
\begin{split}
f(\vecx) &\approx \frac12\left(\oI + (\oI + \oH) + \cdots + (\oH^{M-1} + \oH^{M})\right) 2\overline{u}_D(\vecx)\\
&= \left(\oI + \cdots + \oH^{M-1} + \frac12 \oH^{M}\right) 2\overline{u}_D(\vecx)
\end{split}
\end{equation}
Therefore, the term for the last point $\vecx_M$ should now be multiplied by $1/2$, when compared to just truncating the original series at the $M$-th term. 
For such fixed-length truncation, the resulting estimator is biased, just like MC rendering with a finite path length. 
One could potentially apply the Russian roulette technique to truncate a path without the bias, though we leave it as future work. 

Formally, this transformation of a Neumann series can be mathematically interpreted as an analytic continuation of the series, and many other transformations are possible~\cite{SabelfeldSimonov1994, sabelfeld1991monte}. 
\citet{sabelfeld1991monte} numerically compares some transformations, and our initial experiments also suggest that series acceleration, such as the van Wijngaarden transformation~\cite{van1953transformation}, can reduce the error of the estimator based on Neumann series. 
All the results we show use the modified Neumann series with multiplication of $1/2$ on the last term.

\subsubsection{Derivative and boundary value estimators}
With WoB, in addition to the solution within the domain, we can easily estimate the gradient inside the domain, solution on the boundary, and normal derivative on the boundary, just by replacing the first step of our recursive MC solution estimators. 
For example, we can apply an MC estimator based on the equation for the interior gradient, $\frac{\partial u}{\partial \vecx_k} = -\int \frac{\partial^2 G}{\partial \vecx_k\partial \vecn_\vecy}  \nu \,dA$, instead of \eqref{eq:mc_dirichlet_interior_sol} to get a gradient estimate in the case of the Dirichlet problem estimator with double layer potential formulation. Similar to WoS~\cite{Sawhney:2020:MCG}, we can reuse the same paths to get samples for the solution and the gradient in the interior by changing the initial weight at almost no additional cost.
We however observed increased noise near the boundary in the gradient estimates for this Dirichlet problem estimator. 
This additional noise is likely due to the presence of the hypersingular kernel $\frac{\partial^2 G}{\partial \vecx_k \partial \vecn_\vecy}$ in the computation, which has a very large variance when the interior point $\vecx$ is placed very close to a sampled boundary point $\vecy$. To overcome this issue, we could separate the gradient into the normal and tangential components defined with respect to the nearest boundary point and carefully evaluate each term as described by \citet{SabelfeldSimonov1994}.

The estimates of the normal derivatives on the boundary with WoB use the equations we get by taking the limit of the integral equations for gradient estimation to the boundary. For this Dirichlet estimator, however, the normal derivative estimator derived this way involves a hypersingular integral, which has an infinite variance and cannot be used directly; we will need to transform it to another form for evaluation~\cite{SabelfeldSimonov1994}.

\subsection{Neumann Problems with Direct BIE}\label{sec:NeumannDirect}
The strength of WoB is that we can apply essentially the same approach of building a recursive MC estimator to address other boundary problems than Dirichlet problems. 
While \citet{SabelfeldSimonov1994} proposed using single layer potentials to solve Neumann problems, we propose another formulation based on direct BIEs since this formulation allows us to utilize a different estimator than the one with single layer potentials as we will explain later.
Our formulation uses \eqref{eq:cbie} in combination with the Neumann boundary conditions $\overline{q}_N$ as
\begin{equation}\label{eq:cbie_neumann}
u(\vecx) = -  \int_\Gamma 2\frac{\partial G}{\partial \vecn_\vecy}(\vecx, \vecy) u(\vecy) \,dA_\vecy + \int_\Gamma 2G(\mathbf{x}, \mathbf{y}) \overline{q}_N (\vecy) \,dA_\vecy,
\end{equation}
at a point $\vecx$ on the boundary; the second term can be estimated with another MC estimator without recursion because $\overline{q}_N$ is a known boundary value. 
We have found that it usually suffices to sample one boundary point to estimate the integral of the second term per recursion, though it is possible to have more samples.

The rest is similar to the Dirichlet case described above --- we expand $u$ recursively. We can estimate the unknown $u(\vecx_i)$ as 
\begin{equation}\label{eq:mc_cbie_neumann}
\hat{u}(\vecx_i) \coloneqq -  \frac{2\frac{\partial G}{\partial \vecn_\vecy}(\vecx_i, \vecx_{i+1})}{p_1(\vecx_{i+1}|\vecx_i)} \hat{u}(\vecx_{i+1})  +  \frac{2G(\vecx_i, \vecx'_{i+1})}{p_2(\vecx'_{i+1}|\vecx_i)} \overline{q}_N (\vecx'_{i+1}).
\end{equation}
In general, the two PDFs $p_1$ and $p_2$ can differ, sampling two distinct points $\vecx_{i+1}$ and $\vecx'_{i+1}$ based on the current point $\vecx_i$. This estimator can be used to estimate the interior value based on \eqref{eq:cbie} with 
\begin{equation}\label{eq:mc_cbie_neumann2}
\hat{u}(\vecx_0) \coloneqq -  \frac{\frac{\partial G}{\partial \vecn_\vecy}(\vecx_0, \vecx_1)}{p_1(\vecx_1|\vecx_0)} \hat{u}(\vecx_1)  +  \frac{G(\vecx_0, \vecx'_1)}{p_2(\vecx'_1|\vecx_0)} \overline{q}_N (\vecx'_1),
\end{equation}
where $\vecx_0$ is an interior point and $\vecx_1$ and $\vecx_1'$ are boundary points. 
Estimating the gradient is also possible but with a high variance similar to the case of the Dirichlet problem estimator.

\subsection{Mixed Boundary Problems with Single Layer BIE}\label{sec:mixed_boundary}
For mixed boundary and Robin problems, we adopt a single layer potential formulation (\eqref{eq:indirect_single_u}) where the boundary unknown we need to estimate is $\mu$. This formulation was used for pure Dirichlet, Neumann, and Robin problems by \citet{SabelfeldSimonov1994}, but not for mixed boundary problems.  
This formulation leads to a Fredholm equation of the second kind for parts of the boundary where Neumann or Robin conditions are specified (see Table~\ref{table:equations} for the equations).
However, for parts of the boundary where Dirichlet boundary conditions are specified, this formulation would result in an equation in the form of a Fredholm equation of the \emph{first} kind:
\begin{equation}\label{eq:mixed_first_kind}
     \overline{u}_D(\vecx) = \int_\Gamma G(\vecx, \vecy) \mu(\vecy) \,dA_\vecy \quad\text{for}\quad \vecx\in \Gamma_D.\\
\end{equation}
Unlike the second kind equation, the unknown quantity $\mu$ appears only inside the integral. 
One cannot simply apply recursive MC estimation for equations in the form of the first kind equation since there is no recursion. 
A common approach is to discretize and solve the corresponding matrix equation, which ruins the advantages of WoB over the conventional alternatives. 

Inspired by a similar technique by \citet{SabelfeldSimonov1994}, we propose to transform \eqref{eq:mixed_first_kind} by multiplying by a nonzero constant $k$ on both sides and adding $\mu(\vecx)$ to both sides:
\begin{equation}\label{eq:mixed_second_kind_ish}
    \mu(\vecx) = \mu(\vecx) - k\int_\Gamma G(\vecx, \vecy) \mu(\vecy) \,dA_\vecy +k\overline{u}_D(\vecx) \quad\text{for}\quad \vecx\in \Gamma.
\end{equation}
This equation now has a structure similar to the second kind equation, with the additional $\mu(\vecx)$ on the right-hand side. 
This equation can be estimated recursively, just like the second kind equation. 
We estimate the unknown quantity on the left-hand side by sampling the next point, and estimate the contribution by
\begin{equation}\label{eq:mixed_second_kind_ish_mc}
    \hat{\mu}(\vecx_{i}) \coloneqq \begin{cases} 
        -\frac{1}{p_k}\cdot k\frac{G(\vecx_i, \vecx_{i+1})}{p(\vecx_{i+1}|\vecx_i)} \hat{\mu}(\vecx_{i+1})+k\overline{u}_D(\vecx_i)&\text{with prob. } p_k\\
         \frac{1}{1-p_k}\hat{\mu}(\vecx_{i+1})+k\overline{u}_D(\vecx_i)&\text{with prob. } 1-p_k.
    \end{cases}
\end{equation}
We sample one of the first two terms in \eqref{eq:mixed_second_kind_ish} based on the given probability $p_k$. In the second case when we sample the term $\mu(\vecx)$, we remain at the same point, i.e., $\vecx_{i+1}\coloneqq\vecx_i$, in the next recursion step. 
With this method of handling the first kind equation, we can construct a recursive MC estimator for $\mu$ for mixed boundary problems to recover the solution by another MC integration of \eqref{eq:indirect_single_u}. 

The choice of the multiplication constant $k$ is critical in this estimator, and we picked a value by trial and error.
If it is too small, the bias is higher given the same path length. 
If it is too large, the Neumann series diverges and the estimator fails as justified by \cite{SabelfeldSimonov1994} in the case of pure Dirichlet problems. 
\citet{SabelfeldSimonov1994} also discuss the restriction on the mixture weight $\overline\alpha$ of Robin boundary problems.
We found that deriving theoretical bounds in the case of mixed boundaries are extremely involved and are left for future work.

The gradient estimator with this formulation does not exhibit the additional noise as with the other formulations because this formulation uses an integral kernel with a lower order of singularity. For the normal derivative estimator, we get $\mu$ in two terms, in the integral and outside of the integral. We need to sample either one of the terms or run two paths for the estimates of $\mu$ for the two terms.

One could alternatively employ direct BIEs to get a similar estimator for mixed boundary problems. 
However, we would still encounter a first kind equation, and in practice the resulting schemes are not much different from the indirect BIE-based method above.

\subsection{Sampling Strategies}\label{sec:sampling}
WoB can use various strategies to sample paths as in MC ray tracing, and a well-designed strategy can sharply reduce variance of the estimator. 
Both WoB and MC ray tracing have a vast design space of sampling strategies for different problems, and our proposed strategies in this paper are by no means exhaustive. 
We leave further exploration of different strategies as future work and briefly explain the strategies we implemented in the following. 

\subsubsection{Ray sampling}
The integral kernel $2\frac{\partial G}{\partial \vecn_\vecy}$ is in fact proportional to the differential solid angle of $\vecy$ from another point $\vecx$. 
Similar to the fact that sampling a ray will cancel out the geometry term in rendering, we can use ray tracing from $\vecx$ to perfectly importance sample $2\frac{\partial G}{\partial \vecn_\vecy}$ at any $\vecy$.
The main difference from rendering is that we do not have the visibility term between $\vecy$ and $\vecx$.
We thus have to sample a ray from a sphere, not hemisphere, and sample all the intersection points (not just the first hit) along the ray from $\vecx_i$ via "all-hits" ray intersection queries in general. Such all-hits queries are available and used in ray tracing~\cite{Gribble2014RayTraversal}. %

Due to our recursive formulation, using all hits would cause exponential branching of paths which may not be ideal for GPU ray tracing~\cite{parker2010optix}. 
We can instead pick one intersection out of the $m$ such intersections at random, which results in multiplying the PDF by $1/m$, and the sample contribution is thus multiplied by $m$ but with no branching in recursion. 
We used this approach in all of our results in this paper. 
Since this approach leads to an exponential increase in variance instead of exponential computation cost per sample, we do not claim that this strategy is always better than using all hits and branching exponentially. 

For the special case when the domain is convex, we revert to the original hemispherical sampling strategy with closest-hit query and the PDF in this strategy simplifies to $p(\vecy|\vecx) = -\frac{\partial G}{\partial \vecn_\vecy}(\vecx, \vecy)$ because there is only one hit for a direction within the hemisphere.
The solution estimator for the Dirichlet problem with path length $M$ then becomes 
$
    \hat{u}(\vecx_0) = 2[u_D(\vecx_1) - u_D(\vecx_2) + \cdots] + (-1)^{M} u_D(\vecx_M).
$
This estimator is the one introduced in Section \ref{sec:overview} (and Algorithm \ref{alg:interior_dirichlet_convex}).%

Another strategy is to use only the first hit point, but combine it with direct sampling of a point on the boundary (like sampling a point on light sources in rendering) via multiple importance sampling (MIS)~\cite{veach1995optimally}.
While the first-hit-only strategy would not cover the entire integration domain, MIS ensures an unbiased estimator since direct sampling of a point will cover the entire domain. 
We did not employ this strategy since it often had higher variance in our experiments.  
However, similar to different strategies in MC rendering, there might be certain scenarios where this particular strategy works better than the others. 
We suggest that readers explore different options to determine a suitable strategy for a given problem.

\subsubsection{Backward estimator}
We have so far considered tracing a path by sampling a series of points starting from the evaluation point $\vecx_0$. 
We call this estimator a "backward" estimator, consistent with "backward tracing" in rendering which traces a path of light in a backward manner from the sensor (pixel) all the way to the light source. One can think of our evaluation point as a pixel and a boundary as a light source. 
Note that \citet{qi22bidirectional} adopted the opposite definitions of backward and forward from those in rendering, so their forward estimator corresponds to our backward estimator.

\subsubsection{Forward estimator}
\citet{SabelfeldSimonov1994} proposed an \emph{adjoint estimator} in WoB, which forms paths starting from a point on the boundary. 
We have found that it is analogous to light tracing in rendering, which traces a path starting from a point on the light source and can also be explained via the adjoint of the rendering equation~\cite{christensen2003adjoints}.
We call it a "forward estimator" as in "forward tracing" in rendering. 
In the forward estimator, we will need to make an explicit connection between each point along the path and the evaluation point to compute the contribution of a path to the evaluation point. 
The forward estimator has less variance than the backward estimator in some of the formulations where the integral kernel is proportional to $\frac{\partial G}{\partial \vecn_\vecx}$ as opposed to $\frac{\partial G}{\partial \vecn_\vecy}$. 
In this case, we want to generate a ray from $\vecy$ to sample a point $\vecx$ proportional to its differential solid angle.
For Neumann problems, the formulation with single layer potentials~\cite{SabelfeldSimonov1994} is easier to importance sample by a forward estimator, while our formulation with the direct BIE matches better with a backward estimator.

\subsubsection{Other strategies}\label{sec:othersampling}
Importance sampling only the integral kernel does not perfectly importance sample all the terms, which is also true in rendering. 
For example, the Dirichlet boundary value $\overline{u}_D$ will not be importance sampled that way.
Because our WoB is based on a Fredholm equation of the second kind and we can use ray tracing, it is easy to apply more advanced sampling techniques used in rendering, such as MIS, resampled importance sampling (RIS)~\cite{Talbot2005}, Markov chain MC (MCMC)~\cite{veach1997metropolis, Kelemen2002}, or the zero variance theory~\cite{Krivanek:2014:ZSS}
to implement WoB. 
It contrasts with WoS where applications of these techniques are not necessarily straightforward (e.g., an MIS bidirectional estimator is not available in WoS~\cite{qi22bidirectional}).
We show preliminary results with bidirectional estimators with MIS, resampling via RIS, MCMC WoB, and path reuse as in the virtual point lights method~\cite{keller1997instant}, but many other techniques can be made available to WoB. 
One interesting aspect of WoB that might lead to further development of sampling strategies is that samples can have positive or negative contributions, unlike rendering where all contributions are non-negative.

\subsection{Generalization}

\subsubsection{Exterior problems}\label{sec:exterior}
WoB efficiently handles exterior problems in its basic form.
Instead of the domain $\Omega$, we can solve Laplace's equation in $\mathbb{R}^2\backslash\Omega$ or $\mathbb{R}^3\backslash\Omega$ for exterior problems.
We define the normals to remain oriented outward from the interior domain $\Omega$ and replace the definitions of the boundary values with those obtained by taking the limit from the exterior domain. Moreover, in addition to the boundary conditions, we require that the solution $u(\vecx)$ approaches zero at infinity for exterior problems.
The solvers for the exterior domain largely remain the same as for the interior domain, except for a few sign changes in the terms of the BIEs (see Table~\ref{table:equations}).
Since WoB relies on neither $\epsilon$-shell approximation nor closest point queries in WoS, its accuracy and performance for exterior domains remain the same as for interior problems. 
For example, WoB does not need Kelvin transformations~\cite{Nabizadeh:2021:Kelvin}.

Of the two Dirichlet problem estimators, %
the double layer potential formulation estimator requires additional attention when used for exterior problems.
This formulation allows us to find solutions that decay according to $O(|\vecx|^{1-d})$ as $|\vecx|\rightarrow\infty$, where $d$ is the dimension of the problem. 
It thus cannot handle more general cases where solutions decay as $O(|\vecx|^{2-d})$. 
\citet{SabelfeldSimonov1994} explain how to generalize the double layer potential formulation for such cases. 
The basic idea is to reduce the problem to one without the slowly decaying component with some precomputation before applying WoB.  
We enabled this extension for the scenes in Figures \ref{fig:mc_renderer_result} and \ref{fig:mcmc} where we expect the decay rate $O(|\vecx|^{2-d})$, while we did not do so for the other examples for which we use analytical solutions with decay rate $O(|\vecx|^{1-d})$. 
The single layer potential formulation, on the other hand, supports solutions with a decay rate $O(|\vecx|^{2-d})$ without modification. 
The exterior problem estimators for Neumann, Robin, and mixed boundary problems based on the single layer potential formulation are also general enough to handle the general decay rate of harmonic functions.

\subsubsection{Multiply-connected domain problems}\label{sec:multiply_connected}  We mainly focus on simply connected domains, e.g., domains without holes inside, for simplicity, with the exception of the interior Neumann problem in Fig.~\ref{fig:potential_flow}. 
Multiply connected domains require some additional considerations~\cite{SabelfeldSimonov1994}. 
For Neumann problems with single layer potential, we can apply the same WoB estimators without modifications as long as the standard compatibility condition that the integral of the normal derivative evaluates to zero for each connected domain is satisfied.
For Dirichlet (with double layer potential) and Robin problems (with single layer potential), the situation is more complicated. 
The applicability of WoB in its original form is guaranteed by assuming more artificial compatibility conditions in these cases, and we need to perform some precomputation to modify the problem similarly to the exterior Dirichlet problem estimator. 
We are yet to confirm the applicability and efficiency of these estimators by \citet{SabelfeldSimonov1994} or to derive solvability conditions for Neumann problem estimators based on direct BIE formulation. 
For Dirichlet problems, we observed some successful applications of the single layer potential formulation with multiply-connected domains, but the restriction on the multiplicative constant $k$ seems more strict, and it too requires further investigation.

\subsubsection{Non-zero source term}\label{sec:poisson}
\begin{figure}[t]
\centering
\includegraphics[width=0.19\linewidth]{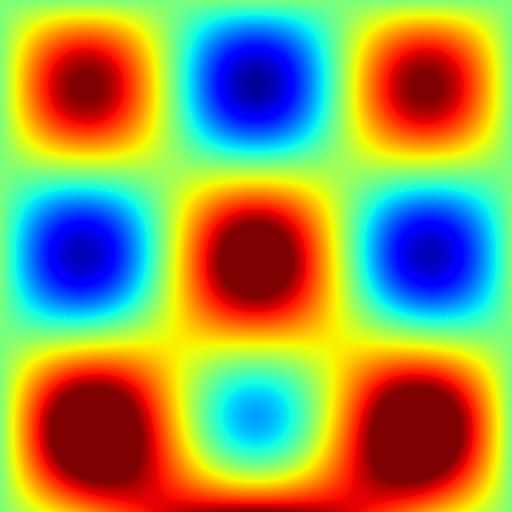}
\includegraphics[width=0.19\linewidth]{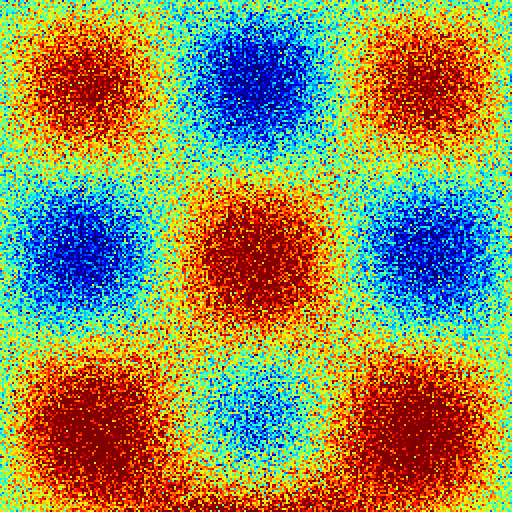}
\includegraphics[width=0.19\linewidth]{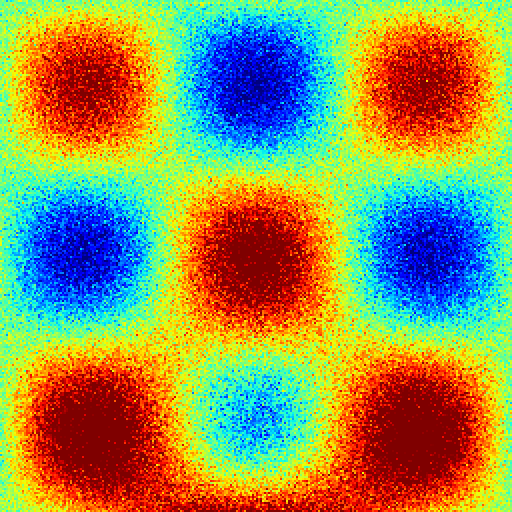}
\includegraphics[width=0.19\linewidth]{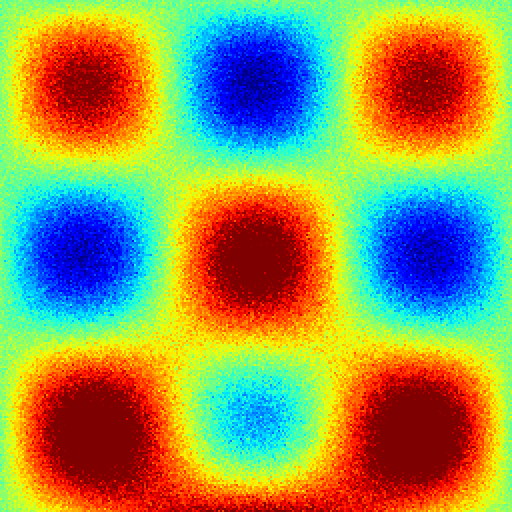}
\includegraphics[width=0.19\linewidth]{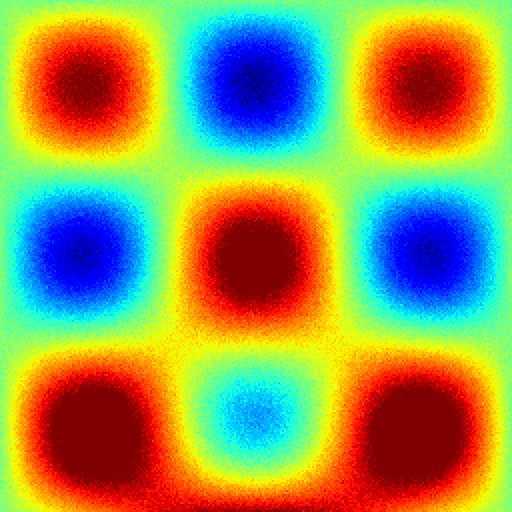}
\includegraphics[width=0.19\linewidth]{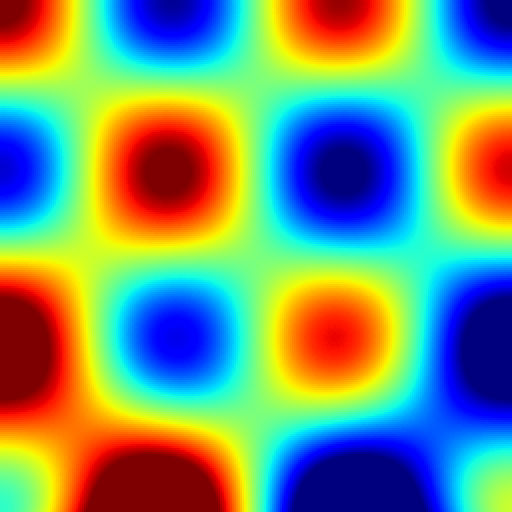}
\includegraphics[width=0.19\linewidth]{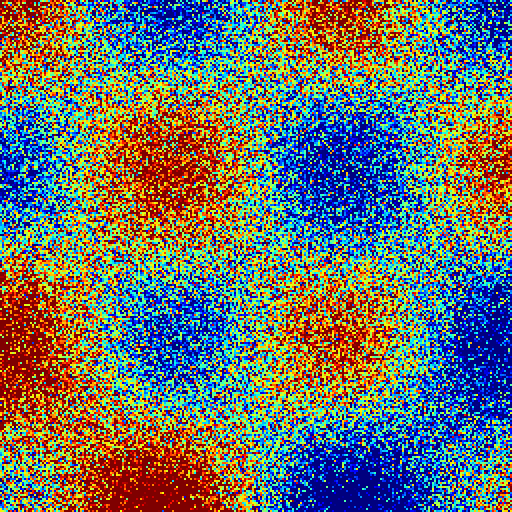}
\includegraphics[width=0.19\linewidth]{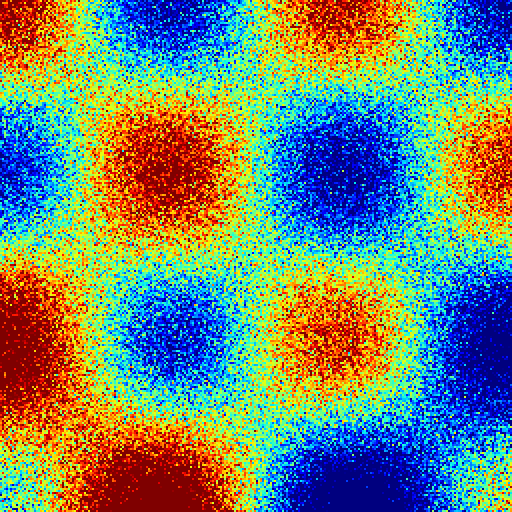}
\includegraphics[width=0.19\linewidth]{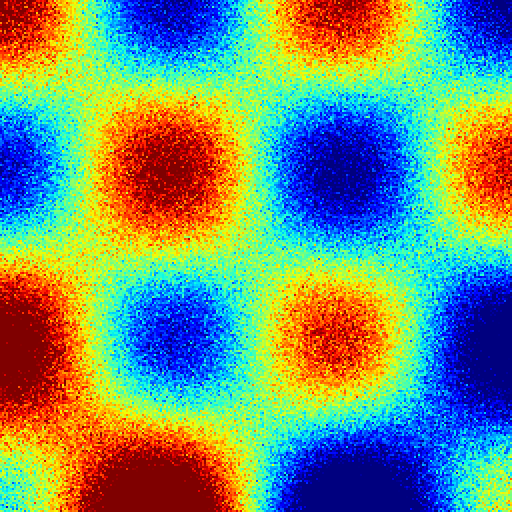}
\includegraphics[width=0.19\linewidth]{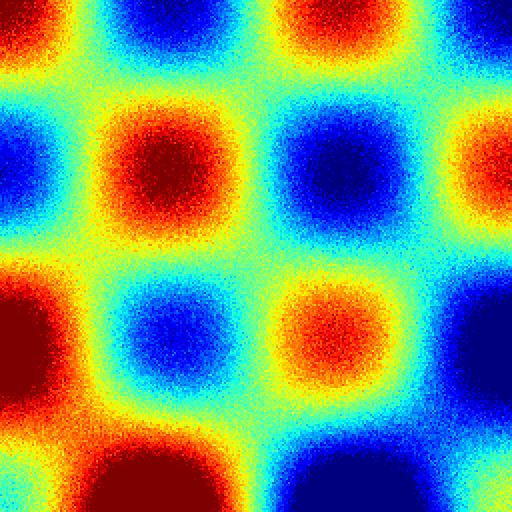}
\begin{minipage}[t]{0.19\linewidth} \centering
\textsf{\small Reference}
\end{minipage}
\begin{minipage}[t]{0.19\linewidth} \centering
\small$N=64$
\end{minipage}
\begin{minipage}[t]{0.19\linewidth} \centering
\small$N=256$
\end{minipage}
\begin{minipage}[t]{0.19\linewidth} \centering
\small$N=1024$
\end{minipage}
\begin{minipage}[t]{0.19\linewidth} \centering
\small$N=4096$
\end{minipage}
\caption{Estimates for Poisson's equation for Dirichlet (top) and Neumann (bottom) problems. WoB can handle the non-zero source term. For each sample path, we used 16 volume samples to estimate all domain integrals.}
\label{fig:poisson_result}
\end{figure}

For Poisson's equation (i.e., $\overline b(\vecx) \neq 0$), we need to additionally sample an interior (or exterior) point to have an MC estimate for the volume integral of $V_0$.
We include this term when we retrieve the boundary values $v(\vecx)$ or $\frac{\partial v}{\partial \vecn}(\vecx)$ in Laplace's equation and when we compute the function $V_0$ in the relation $u(\vecx) = v(\vecx) - V_0(\vecx)$. 
One strategy is to sample such a point uniformly within the domain, but a well-designed sampling strategy can reduce variance. 
For interior problems, one possible sampling strategy is to sample a point along a line that goes through the point $\vecx$. %
Another possibility is to sample the point depending on the distribution of the source term.
Fig~\ref{fig:poisson_result} shows examples for Dirichlet and Neumann problems by sampling interior points uniformly in terms of the area measure. 
We focus on Laplace's equation in the other results since the volume integral of $V_0$ can always be added trivially.
\citet{Sawhney:2022:GFMC} dismissed WoB as being formulated only for Laplace's equation, but as we show, WoB is certainly capable of handling Poisson's equation. %

\begin{figure*}[t]
\centering
\begin{minipage}[b]{0.015\linewidth} \centering
\begin{sideways}
\textbf{Interior}
\end{sideways}
\end{minipage}
\includegraphics[width=0.153\linewidth]{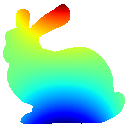}
\includegraphics[width=0.153\linewidth]{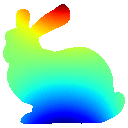}
\includegraphics[width=0.153\linewidth]{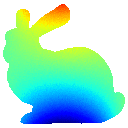}
\includegraphics[width=0.153\linewidth]{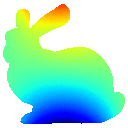}
\includegraphics[width=0.153\linewidth]{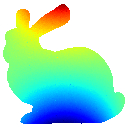}
\includegraphics[width=0.153\linewidth]{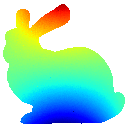}
\includegraphics[width=0.0119\linewidth]{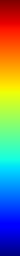}
\begin{minipage}[b]{0.015\linewidth} \centering
\begin{sideways}
-2.04\;\;\;\;\;\;\;0\;\;\;\;\;\;\;\;\;2.04
\end{sideways}
\end{minipage}

\centering
\begin{minipage}[b]{0.015\linewidth}\;\end{minipage}
\begin{minipage}[b]{0.153\linewidth}\;\end{minipage}
\resizebox{0.153\linewidth}{!}{\includegraphics{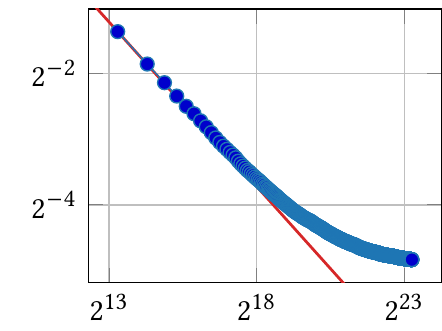}}
\resizebox{0.153\linewidth}{!}{\includegraphics{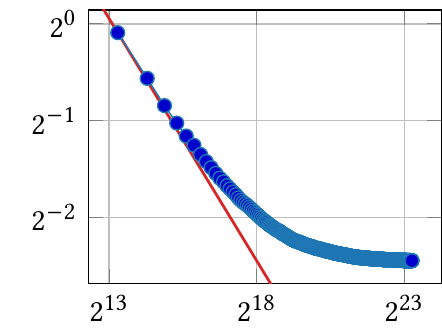}}
\resizebox{0.153\linewidth}{!}{\includegraphics{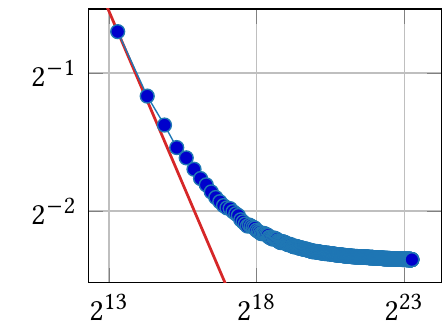}}
\resizebox{0.153\linewidth}{!}{\includegraphics{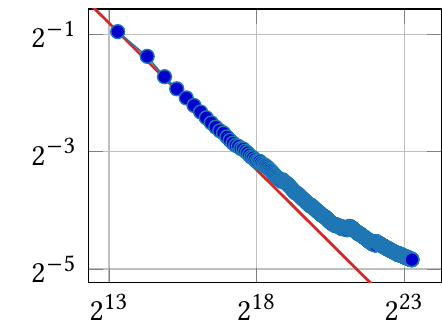}}
\resizebox{0.153\linewidth}{!}{\includegraphics{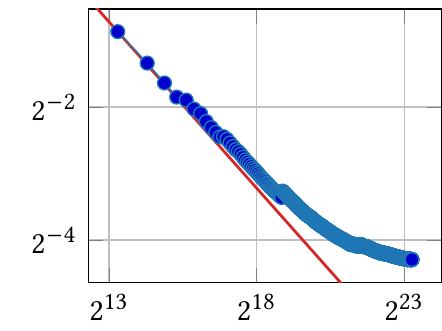}}
\begin{minipage}[b]{0.0119\linewidth}\;\end{minipage}
\begin{minipage}[b]{0.015\linewidth}\;\end{minipage}

\centering
\begin{minipage}[b]{0.015\linewidth} \centering
\begin{sideways}
\textbf{ Exterior}
\end{sideways}
\end{minipage}
\includegraphics[width=0.153\linewidth]{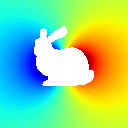}
\includegraphics[width=0.153\linewidth]{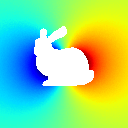}
\includegraphics[width=0.153\linewidth]{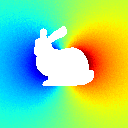}
\includegraphics[width=0.153\linewidth]{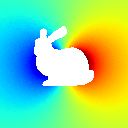}
\includegraphics[width=0.153\linewidth]{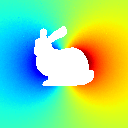}
\includegraphics[width=0.153\linewidth]{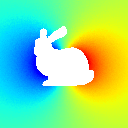}
\includegraphics[width=0.0119\linewidth]{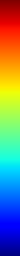}
\begin{minipage}[b]{0.015\linewidth} \centering
\begin{sideways}
-5.25\;\;\;\;\;\;\;0\;\;\;\;\;\;\;\;\;5.25
\end{sideways}
\end{minipage}

\centering
\begin{minipage}[b]{0.015\linewidth}\;\end{minipage}
\begin{minipage}[b]{0.153\linewidth} \centering \;\end{minipage}
\resizebox{0.153\linewidth}{!}{\includegraphics{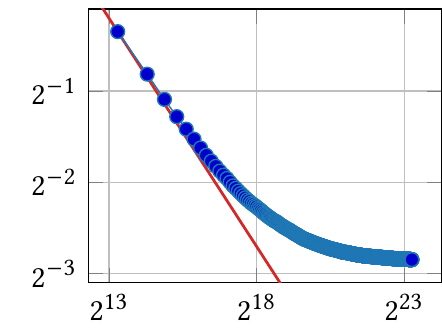}}
\resizebox{0.153\linewidth}{!}{\includegraphics{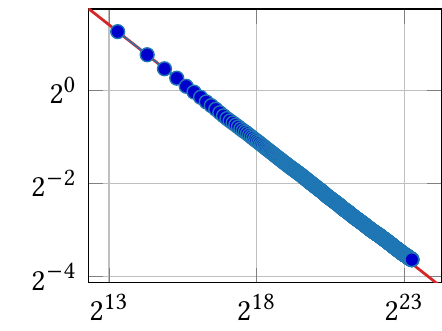}}
\resizebox{0.153\linewidth}{!}{\includegraphics{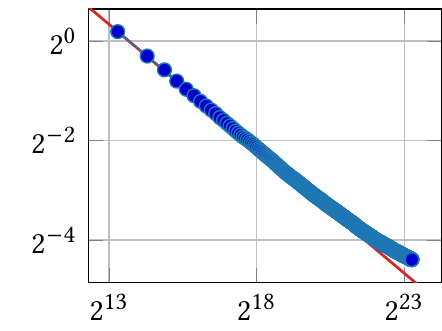}}
\resizebox{0.153\linewidth}{!}{\includegraphics{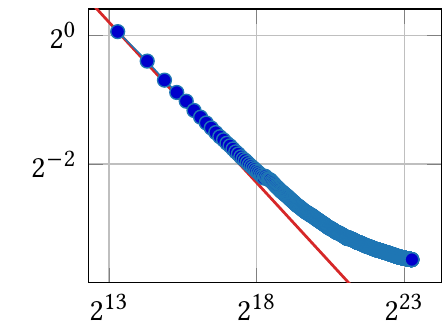}}
\resizebox{0.153\linewidth}{!}{\includegraphics{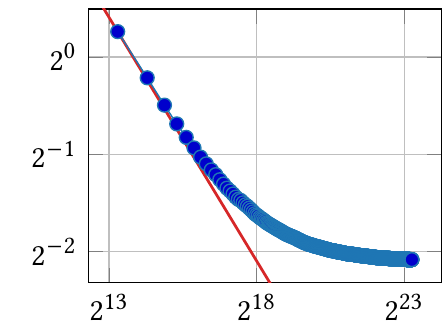}}
\begin{minipage}[b]{0.0119\linewidth}\;\end{minipage}
\begin{minipage}[b]{0.015\linewidth}\;\end{minipage}

\begin{minipage}[b]{0.015\linewidth} \centering
\;
\end{minipage}
\begin{minipage}[t]{0.153\linewidth} \centering
\;\\ %
\textbf{BIE Formulation}\\
\textbf{Sampling Direction}\\
\textbf{Sampling Method}
\end{minipage}
\begin{minipage}[t]{0.153\linewidth} \centering
\textbf{Dirichlet}\\
{double layer}\\
{backward}\\
{ray intersection}
\end{minipage}
\begin{minipage}[t]{0.153\linewidth} \centering
\textbf{Neumann}\\
{direct}\\
{backward}\\
{ray intersection}
\end{minipage}
\begin{minipage}[t]{0.153\linewidth} \centering
\textbf{Neumann}\\
{single layer}\\
{forward}\\
{ray intersection}
\end{minipage}
\begin{minipage}[t]{0.153\linewidth} \centering
\textbf{Robin}\\
{single layer}\\
{backward}\\
{RIS}
\end{minipage}
\begin{minipage}[t]{0.153\linewidth} \centering
\textbf{Mixed}\\
{single layer}\\
{backward}\\
{RIS}
\end{minipage}
\begin{minipage}[b]{0.0119\linewidth}\;\end{minipage}
\begin{minipage}[b]{0.015\linewidth}\;\end{minipage}

\caption{WoB applied to various problems. For the interior (top) and exterior (bottom) problems, we run WoB with path length $M=4$ and $N=10^7$ samples per evaluation point with the formulations and sampling techniques as labeled. The solution estimates are visualized with a color map. For each problem, we show the absolute root mean square error (vertical axis, varying scales) with respect to the number of samples (horizontal axis) measured against the reference analytical solution (left). 
The red lines show the $\mathcal{O}(1/\sqrt{N})$ decay rate for reference.
}
\label{fig:2d_results}
\end{figure*}

\begin{figure*}
\centering

\raisebox{-0.5\height}[0pt][0pt]{
\begin{minipage}[b]{0.015\linewidth} \centering
\begin{sideways}
\textbf{Neumann - Direct}
\end{sideways}
\end{minipage}
}
\includegraphics[width=0.153\linewidth]{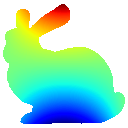}
\includegraphics[width=0.153\linewidth]{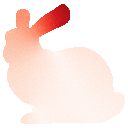}
\includegraphics[width=0.153\linewidth]{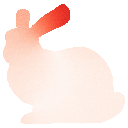}
\includegraphics[width=0.153\linewidth]{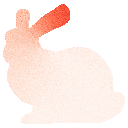}
\includegraphics[width=0.153\linewidth]{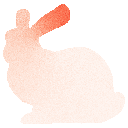}
\includegraphics[width=0.153\linewidth]{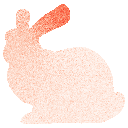}
\begin{minipage}[b]{0.0119\linewidth} \centering \includegraphics[width=\linewidth]{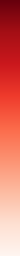}\end{minipage}
\begin{minipage}[b]{0.015\linewidth} \centering
\begin{sideways}
0\;\;\;\;\;\;\;\;\;\;\;\;\;\;\;\;\;\;\;\;\;\;\;\;\;\;\;\;1
\end{sideways}
\end{minipage}

\centering
\begin{minipage}[b]{0.015\linewidth}\;\end{minipage}
\begin{minipage}[b]{0.153\linewidth}\;\end{minipage}
\resizebox{0.153\linewidth}{!}{\includegraphics{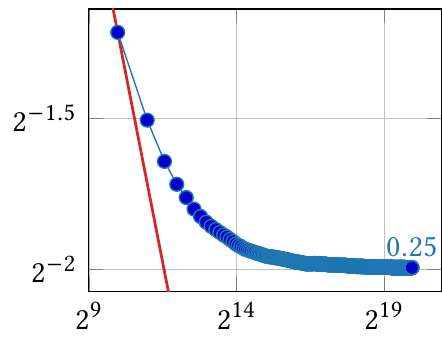}}
\resizebox{0.153\linewidth}{!}{\includegraphics{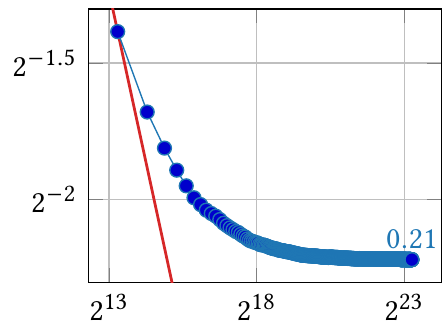}}
\resizebox{0.153\linewidth}{!}{\includegraphics{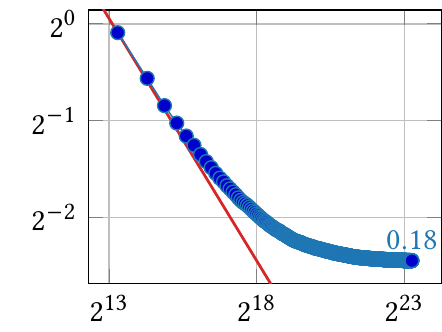}}
\resizebox{0.153\linewidth}{!}{\includegraphics{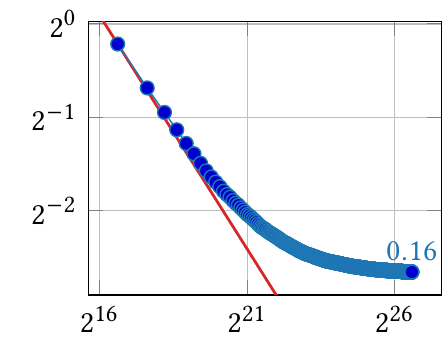}}
\resizebox{0.153\linewidth}{!}{\includegraphics{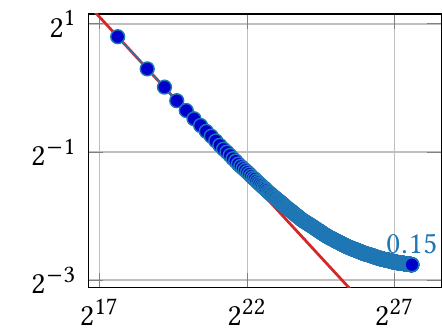}}
\begin{minipage}[b]{0.0119\linewidth}\;\end{minipage}
\begin{minipage}[b]{0.015\linewidth}\;\end{minipage}

\raisebox{-0.5\height}[0pt][0pt]{
\begin{minipage}[b]{0.015\linewidth} \centering
\begin{sideways}
\textbf{Neumann - Single Layer}
\end{sideways}
\end{minipage}
}
\includegraphics[width=0.153\linewidth]{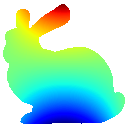}
\includegraphics[width=0.153\linewidth]{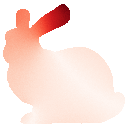}
\includegraphics[width=0.153\linewidth]{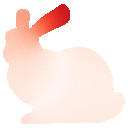}
\includegraphics[width=0.153\linewidth]{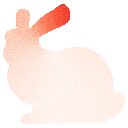}
\includegraphics[width=0.153\linewidth]{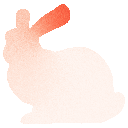}
\includegraphics[width=0.153\linewidth]{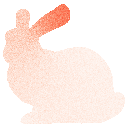}
\begin{minipage}[b]{0.0119\linewidth} \centering \includegraphics[width=\linewidth]{Figures/PathLengths/error_legend.png}\end{minipage}
\begin{minipage}[b]{0.015\linewidth} \centering
\begin{sideways}
0\;\;\;\;\;\;\;\;\;\;\;\;\;\;\;\;\;\;\;\;\;\;\;\;\;\;\;\;1
\end{sideways}
\end{minipage}

\centering
\begin{minipage}[b]{0.015\linewidth}\;\end{minipage}
\begin{minipage}[b]{0.153\linewidth}\;\end{minipage}
\resizebox{0.153\linewidth}{!}{\includegraphics{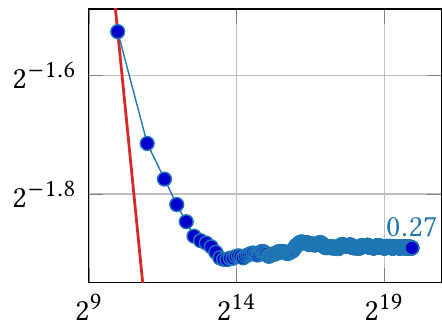}}
\resizebox{0.153\linewidth}{!}{\includegraphics{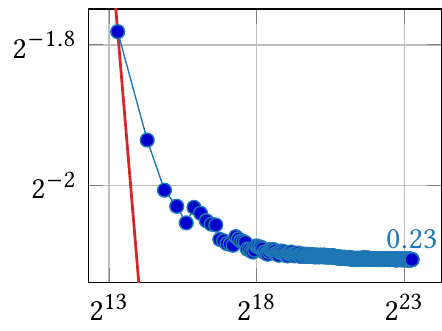}}
\resizebox{0.153\linewidth}{!}{\includegraphics{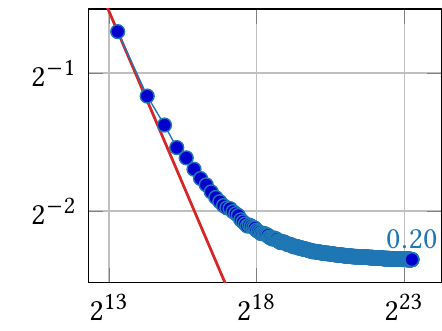}}
\resizebox{0.153\linewidth}{!}{\includegraphics{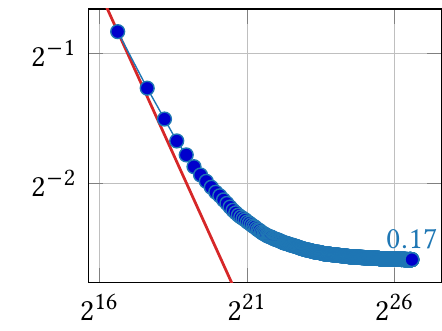}}
\resizebox{0.153\linewidth}{!}{\includegraphics{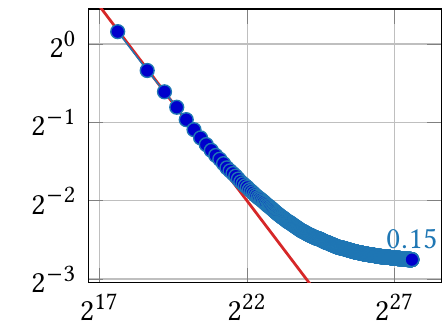}}
\begin{minipage}[b]{0.0119\linewidth}\;\end{minipage}
\begin{minipage}[b]{0.015\linewidth}\;\end{minipage}

\raisebox{-0.5\height}[0pt][0pt]{
\begin{minipage}[b]{0.015\linewidth} \centering
\begin{sideways}
\textbf{Dirichlet}
\end{sideways}
\end{minipage}
}
\includegraphics[width=0.153\linewidth]{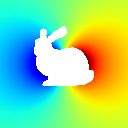}
\includegraphics[width=0.153\linewidth]{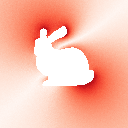}
\includegraphics[width=0.153\linewidth]{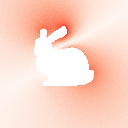}
\includegraphics[width=0.153\linewidth]{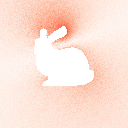}
\includegraphics[width=0.153\linewidth]{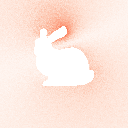}
\includegraphics[width=0.153\linewidth]{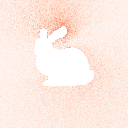}
\begin{minipage}[b]{0.0119\linewidth} \centering \includegraphics[width=\linewidth]{Figures/PathLengths/error_legend.png}\end{minipage}
\begin{minipage}[b]{0.015\linewidth} \centering
\begin{sideways}
0\;\;\;\;\;\;\;\;\;\;\;\;\;\;\;\;\;\;\;\;\;\;\;\;\;\;\;\;1
\end{sideways}
\end{minipage}

\centering
\begin{minipage}[b]{0.015\linewidth}\;\end{minipage}
\begin{minipage}[b]{0.153\linewidth}\;\end{minipage}
\resizebox{0.153\linewidth}{!}{\includegraphics{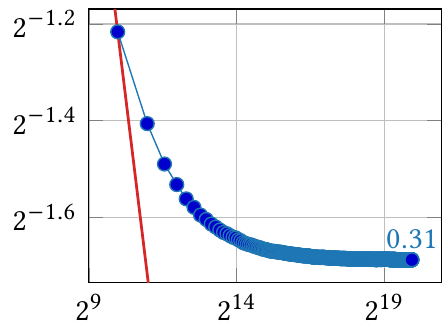}}
\resizebox{0.153\linewidth}{!}{\includegraphics{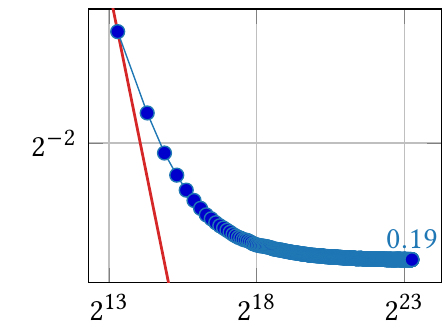}}
\resizebox{0.153\linewidth}{!}{\includegraphics{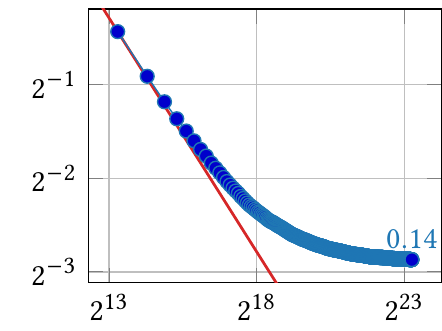}}
\resizebox{0.153\linewidth}{!}{\includegraphics{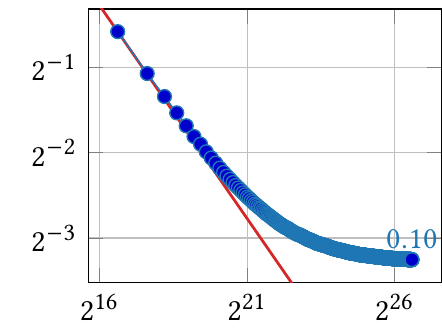}}
\resizebox{0.153\linewidth}{!}{\includegraphics{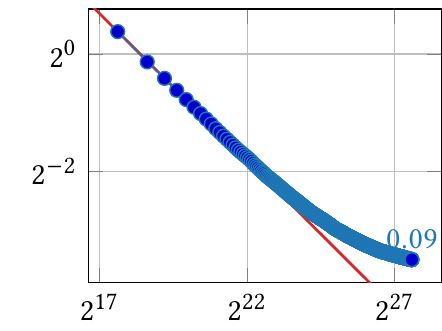}}
\begin{minipage}[b]{0.0119\linewidth}\;\end{minipage}
\begin{minipage}[b]{0.015\linewidth}\;\end{minipage}

\raisebox{-0.5\height}[0pt][0pt]{
\begin{minipage}[b]{0.015\linewidth} \centering
\begin{sideways}
\textbf{Mixed}
\end{sideways}
\end{minipage}
}
\includegraphics[width=0.153\linewidth]{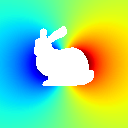}
\includegraphics[width=0.153\linewidth]{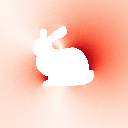}
\includegraphics[width=0.153\linewidth]{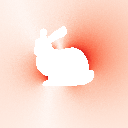}
\includegraphics[width=0.153\linewidth]{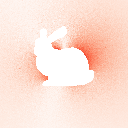}
\includegraphics[width=0.153\linewidth]{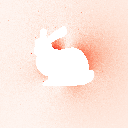}
\includegraphics[width=0.153\linewidth]{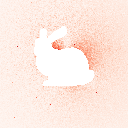}
\begin{minipage}[b]{0.0119\linewidth} \centering \includegraphics[width=\linewidth]{Figures/PathLengths/error_legend.png}\end{minipage}
\begin{minipage}[b]{0.015\linewidth} \centering
\begin{sideways}
0\;\;\;\;\;\;\;\;\;\;\;\;\;\;\;\;\;\;\;\;\;\;\;\;\;1.5
\end{sideways}
\end{minipage}

\centering
\begin{minipage}[b]{0.015\linewidth}\;\end{minipage}
\begin{minipage}[b]{0.153\linewidth}\;\end{minipage}
\resizebox{0.153\linewidth}{!}{\includegraphics{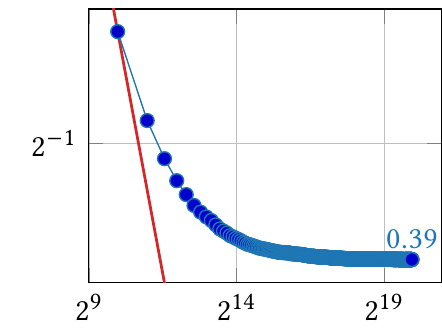}}
\resizebox{0.153\linewidth}{!}{\includegraphics{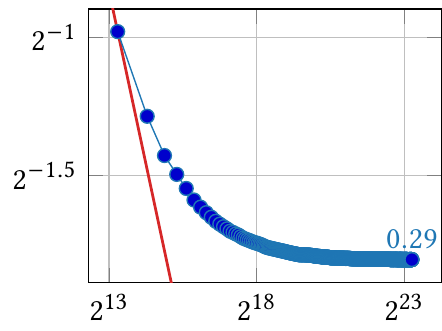}}
\resizebox{0.153\linewidth}{!}{\includegraphics{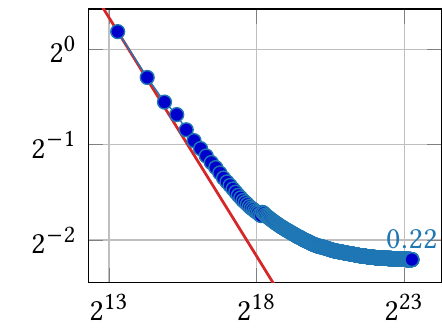}}
\resizebox{0.153\linewidth}{!}{\includegraphics{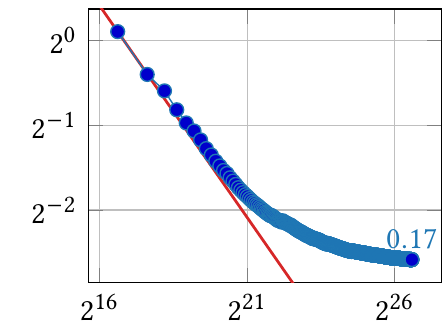}}
\resizebox{0.153\linewidth}{!}{\includegraphics{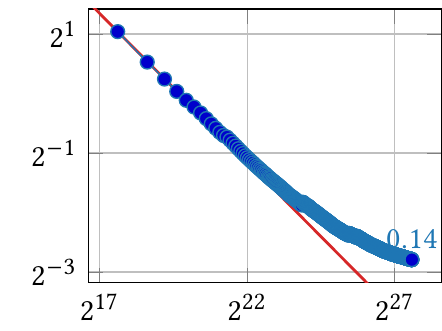}}
\begin{minipage}[b]{0.0119\linewidth}\;\end{minipage}
\begin{minipage}[b]{0.015\linewidth}\;\end{minipage}

\begin{minipage}[b]{0.015\linewidth} \centering
\;
\end{minipage}
\begin{minipage}[t]{0.153\linewidth} \centering
\small\textsf{Analytical Solution}
\end{minipage}
\begin{minipage}[t]{0.153\linewidth} \centering
$M=2$
\end{minipage}
\begin{minipage}[t]{0.153\linewidth} \centering
$M=3$
\end{minipage}
\begin{minipage}[t]{0.153\linewidth} \centering
$M=4$
\end{minipage}
\begin{minipage}[t]{0.153\linewidth} \centering
$M=5$
\end{minipage}
\begin{minipage}[t]{0.153\linewidth} \centering
$M=6$
\end{minipage}
\begin{minipage}[t]{0.0119\linewidth} \centering
\;
\end{minipage}
\begin{minipage}[t]{0.015\linewidth} \centering
\;
\end{minipage}
\caption{Path truncation error study. For the interior Neumann problem estimators and the exterior Dirichlet and mixed boundary problem estimators in Fig.~\ref{fig:2d_results}, we show the absolute errors with different path lengths $M$. We show the absolute root mean square error (vertical axis) with respect to the number of samples (horizontal axis), and also show the remaining error in blue text. As path length increases the bias decreases, but larger numbers of samples are needed to achieve convergence.}
\vspace{-1EM}
\label{fig:path_truncation}
\end{figure*}

\begin{figure*}[t]
    \includegraphics[width=0.16\linewidth]{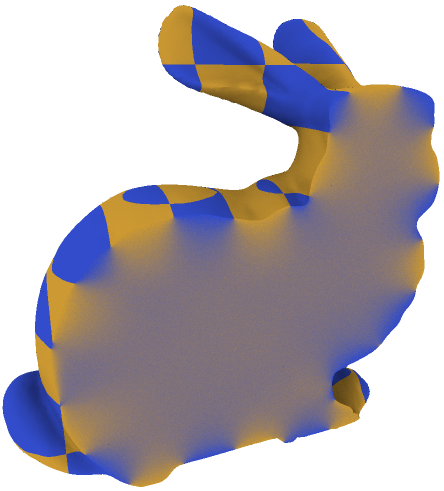}
    \includegraphics[width=0.16\linewidth]{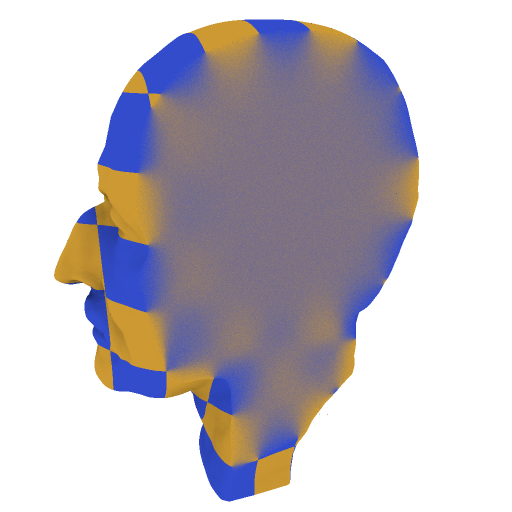}
    \includegraphics[width=0.16\linewidth]{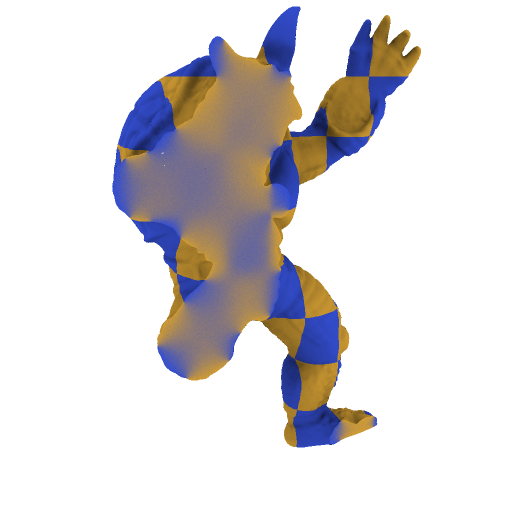}
    \includegraphics[width=0.16\linewidth]{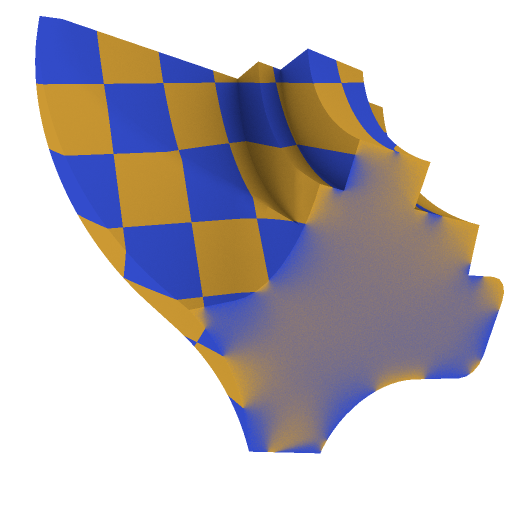}
    \includegraphics[width=0.16\linewidth]{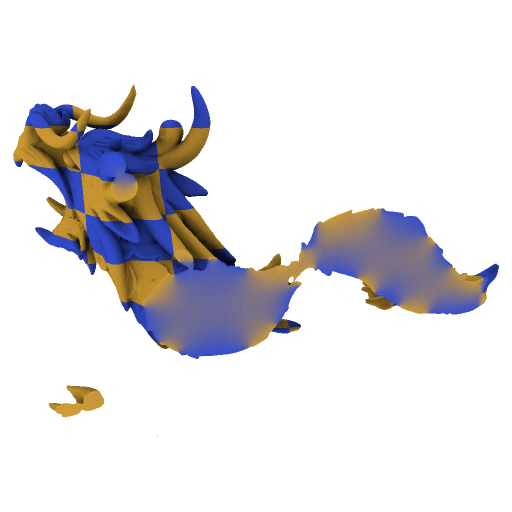}
    \includegraphics[width=0.16\linewidth]{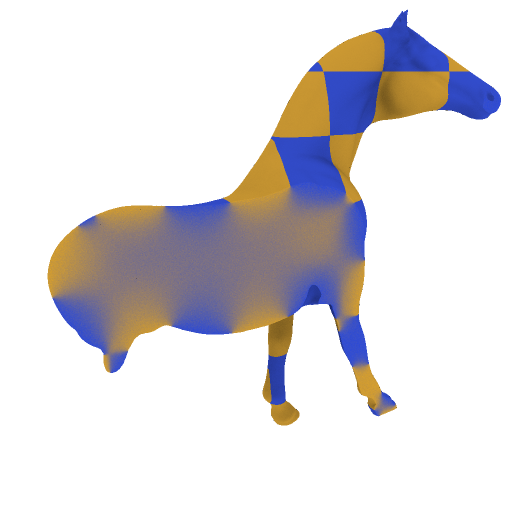}
    \includegraphics[width=0.16\linewidth]{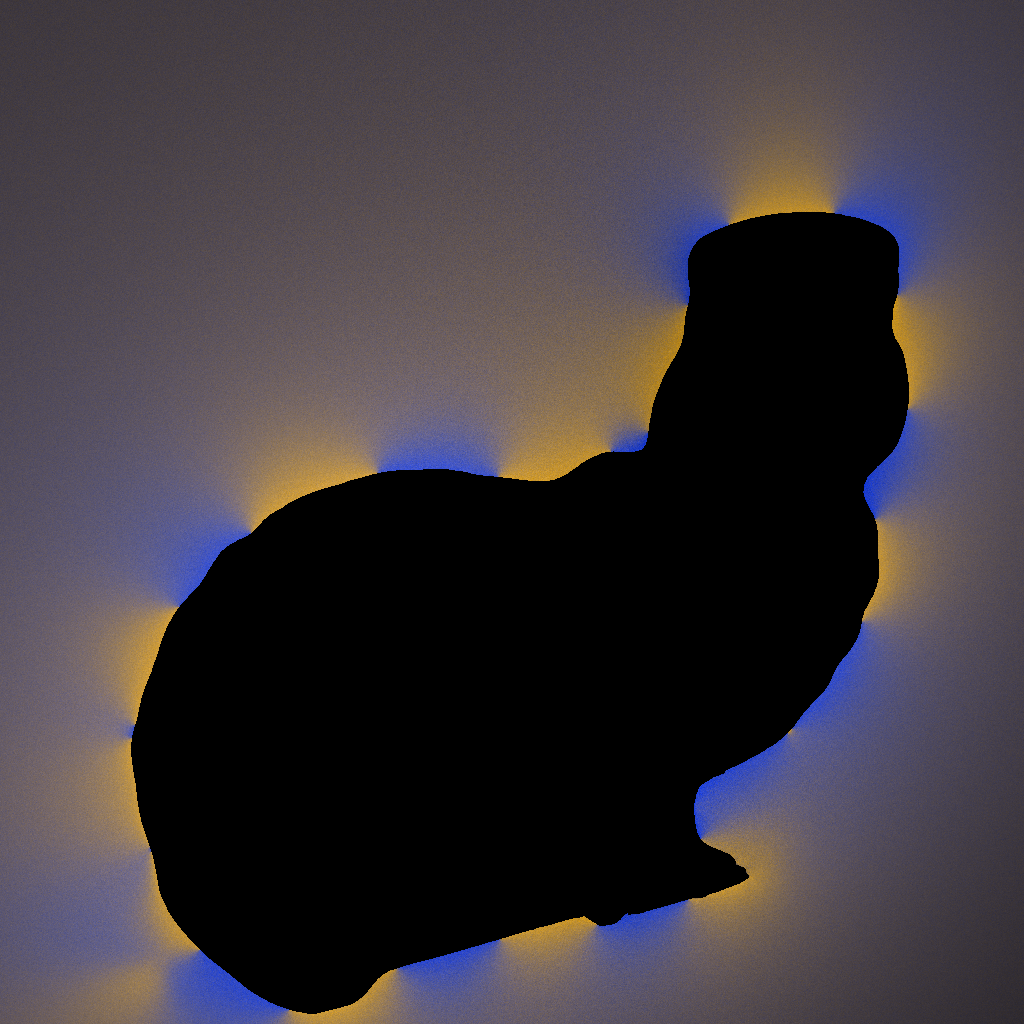}
    \includegraphics[width=0.16\linewidth]{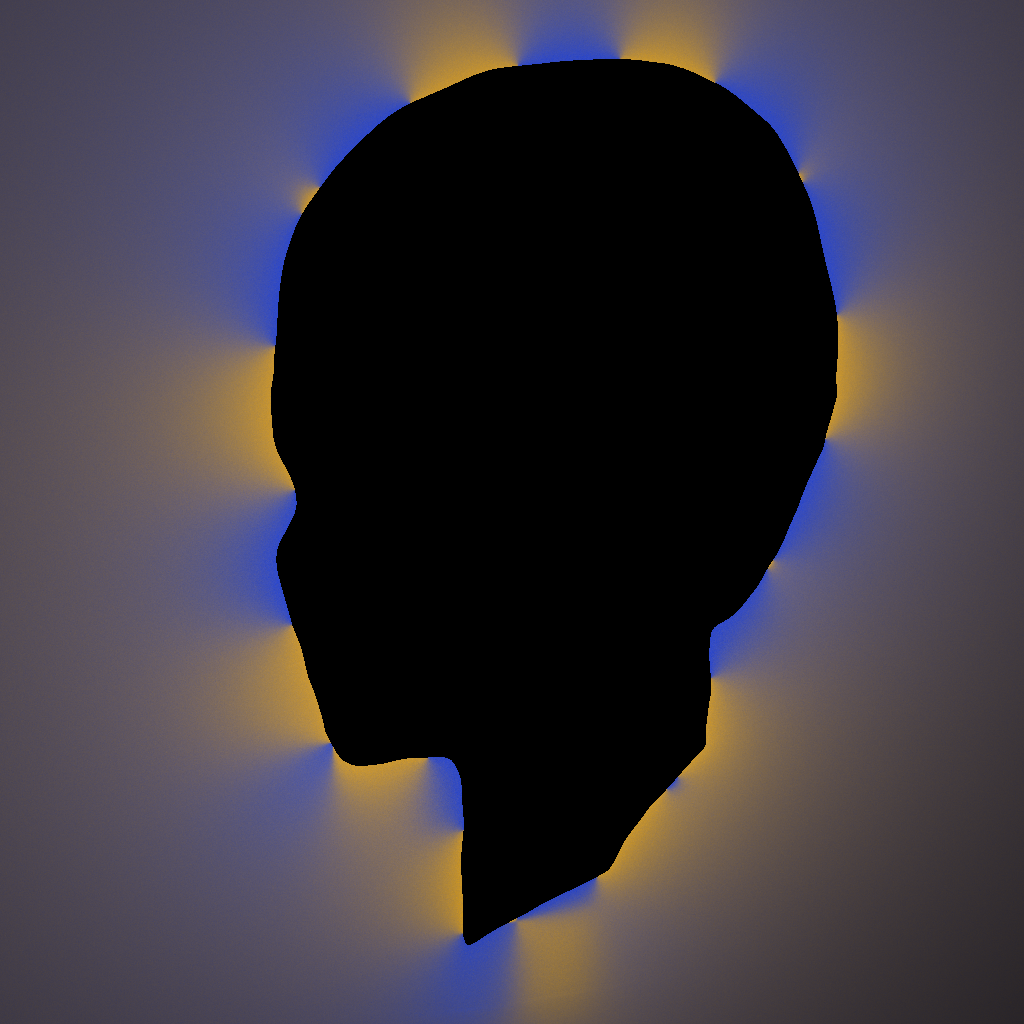}
    \includegraphics[width=0.16\linewidth]{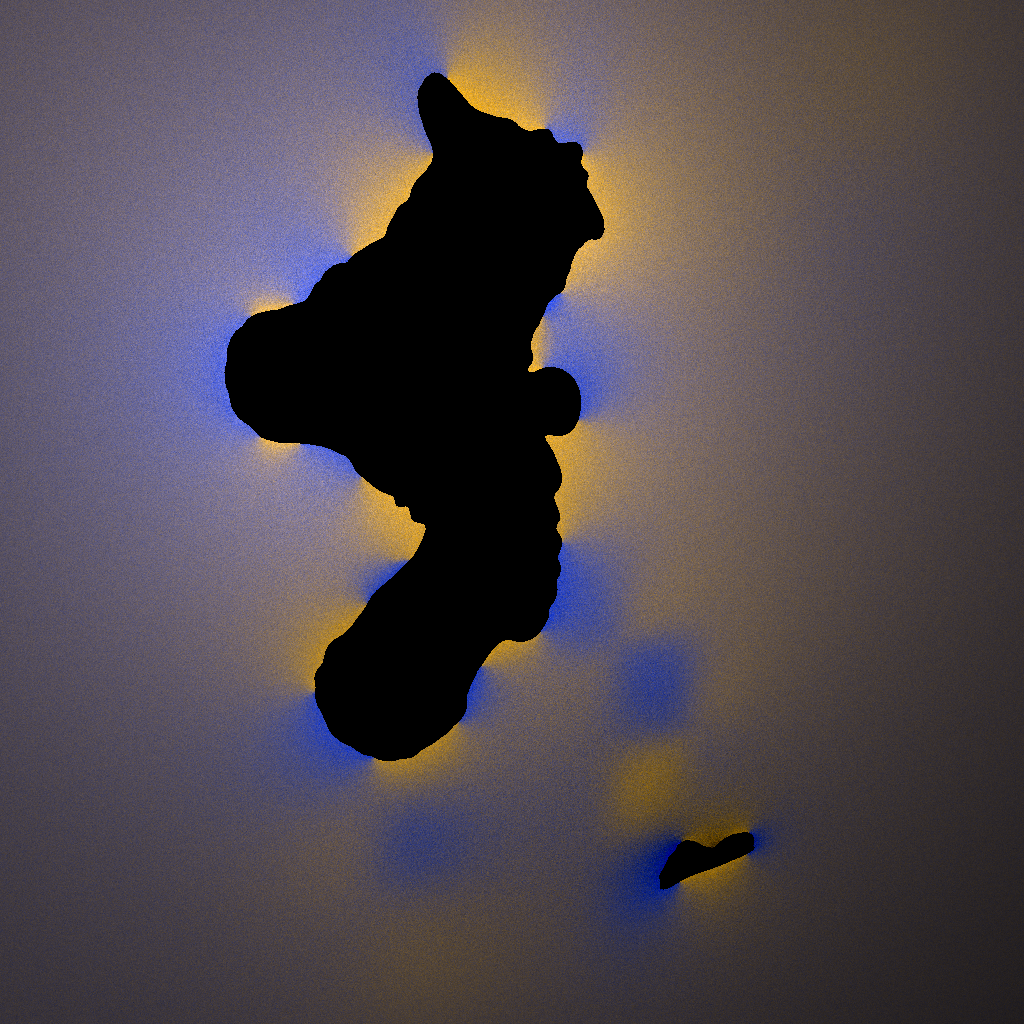}
    \includegraphics[width=0.16\linewidth]{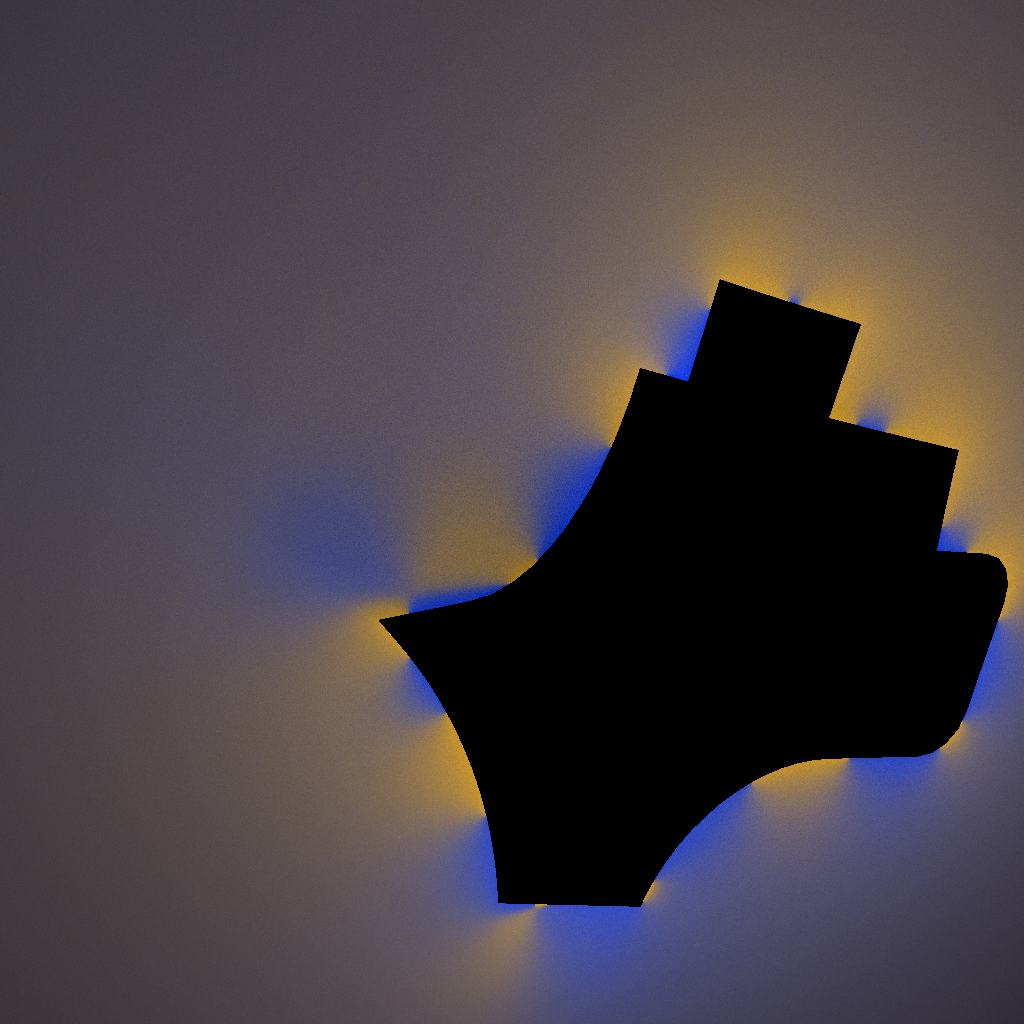}
    \includegraphics[width=0.16\linewidth]{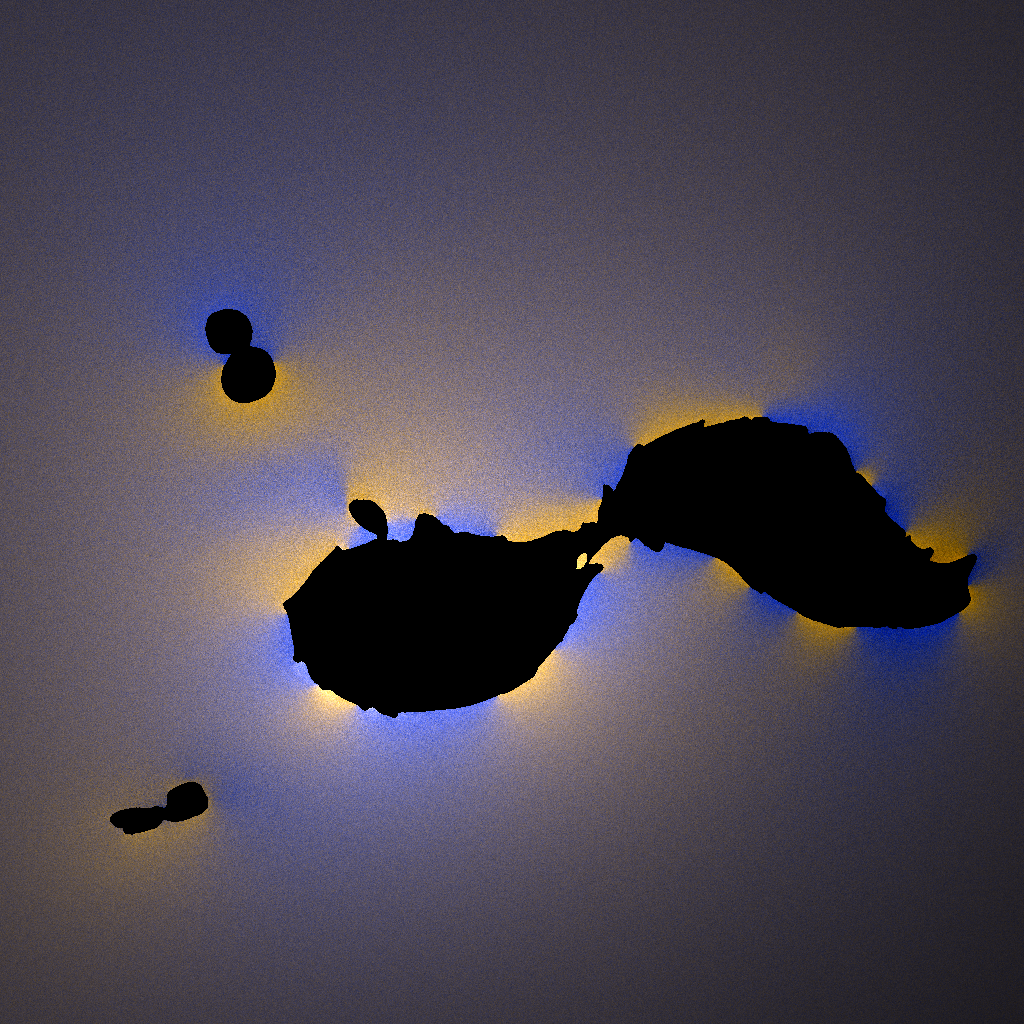}
    \includegraphics[width=0.16\linewidth]{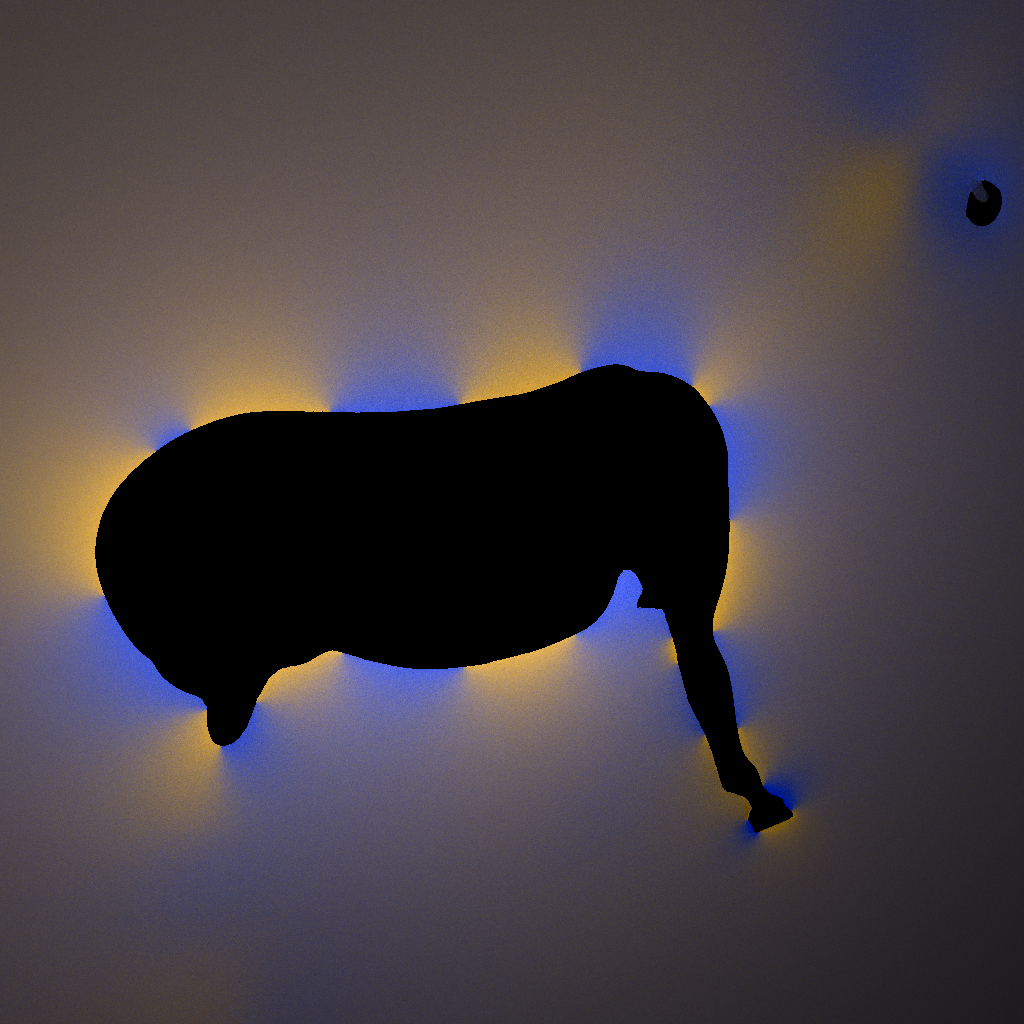}
    \caption{Results of a WoB solver implemented on top of an MC ray tracing system. WoB's strong similarity to MC ray tracing makes such an implementation easy to carry out. The images show the estimated solution for interior (top) and exterior (bottom) Dirichlet problems on a cutting plane. WoB can solve both problems efficiently with a unified MC ray tracing solver. Ambient occlusion was computed at the same time as the solution using the same rendering system.}
    \label{fig:mc_renderer_result}
\end{figure*}

\section{Results}

\paragraph{Generality of WoB}
The same WoB framework based on boundary integral equations successfully handles Dirichlet, Neumann, Robin, and mixed boundary problems, for either interior or exterior domains. Figures \ref{fig:3d_teaser} and \ref{fig:2d_results} show the results of WoB for interior and exterior Laplace problems with known analytical solutions, where we set the boundary conditions to satisfy the known analytical solution. For Robin boundaries, we use a constant value $\overline\alpha =1$ for the mix weight in these examples. For the mixed boundary problems, we uniformly randomly assign one of the Dirichlet, Neumann, or Robin boundary conditions to each boundary triangle or line segment, and use $k=4$ and $p_k =2/3$.
We show results using the sampling strategies described earlier.
For Robin and mixed boundary problems, a pure ray sampling strategy without MIS would miss nonzero contributions, so we instead use RIS with candidates generated uniformly over the boundary to approximately sample the integral kernel. In Fig.~\ref{fig:2d_results}, the number of candidates is 16.
All results in Fig.~\ref{fig:2d_results} are generated with path length $M=4$ and sample path count per evaluation point $N=10^7$ for consistency, using a highly parallelized CUDA implementation.
We observe that \emph{all} of the results are consistent with the analytical solution as expected. %

\paragraph{Convergence rate and truncation error}
WoB exhibits the expected MC convergence rate of $\mathcal{O}(1/\sqrt{N})$. %
The RMSE curve can eventually become flat when we take enough samples because of the error introduced by path truncation. 
This error decreases as we increase the path length although that introduces larger variance to the estimator. 
For the results in Fig.~\ref{fig:2d_results}, we observe in particular a relatively large truncation error for the two interior Neumann problem estimators and the exterior Dirichlet and mixed boundary problem estimators. We thus show their results with different path lengths in Fig.~\ref{fig:path_truncation}. As expected, we observe that having longer paths decreases the truncation error at the cost of having a higher variance, requiring a larger sample count to converge. 
It might be possible to apply stochastic truncation as in MC rendering~\cite{misso2022unbiased} to avoid this error, though a careful investigation is needed to handle subtle differences between WoB and MC rendering.

\paragraph{WoB within MC rendering} Fig.~\ref{fig:mc_renderer_result} show the results of WoB implemented within our MC ray tracing system. 
Our implementation of WoB utilizes the existing functionalities of MC ray tracing such as ray-object intersection, stochastic sampling, and textures (for boundary values). 
The code for WoB itself is roughly 100 lines and can support both interior and exterior problems seamlessly. 
Visualization was done by simultaneously running rendering with ambient occlusion using the MC ray tracing system.  
\begin{wrapfigure}{r}{0.3\linewidth}
   \includegraphics[width=\linewidth]{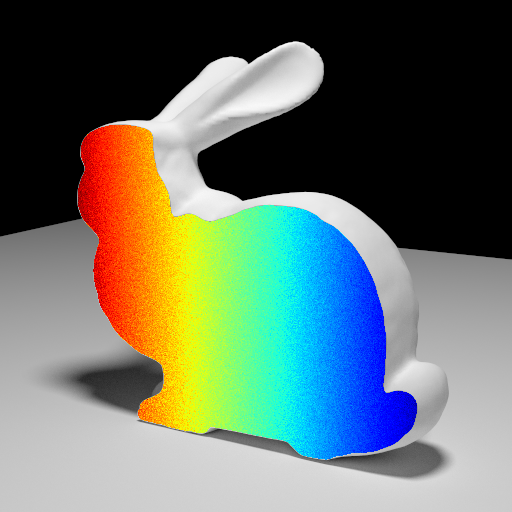}
\end{wrapfigure}
The inset figure shows a result generated with our prototype interior Dirichlet solver on top of a popular open-source renderer for research purposes, PBRT~\cite{Pharr:2018:PBRT}, with minimal modifications. 
For this problem, we set the boundary values such that we expect to see a linear horizontal gradient of solution values mapped to colors ranging from red to green to blue.

\begin{figure}[t]
\centering
\includegraphics[width=0.24\linewidth]{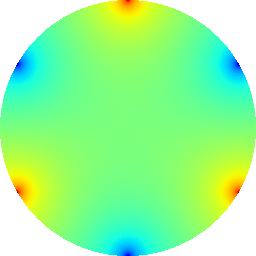}
\includegraphics[width=0.24\linewidth]{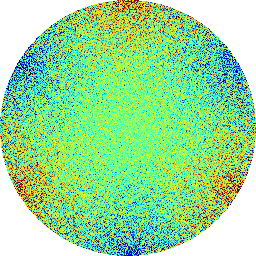}
\includegraphics[width=0.24\linewidth]{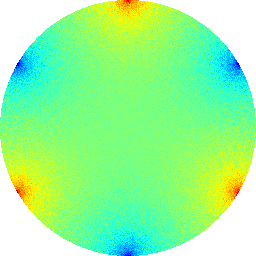}
\includegraphics[width=0.24\linewidth]{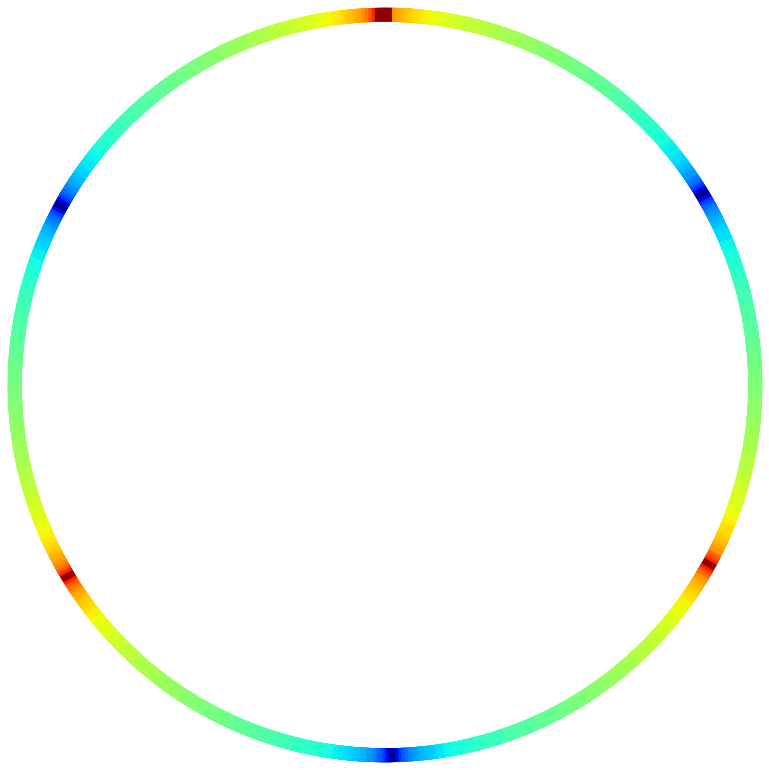}
\begin{minipage}[t]{0.24\linewidth} \centering
\textsf{\small(a) reference}
\end{minipage}
\begin{minipage}[t]{0.24\linewidth} \centering
\textsf{\small(b) uniform sampling}
\end{minipage}
\begin{minipage}[t]{0.24\linewidth} \centering
\textsf{\small(c) boundary sampling}
\end{minipage}
\begin{minipage}[t]{0.24\linewidth} \centering
\textsf{\small(d) estimate on boundary}
\end{minipage}
\caption{Neumann solver with single layer formulation. The boundary value $\partial u/\partial \vecn=0$ except for around the six points. By sampling the start of paths according to the distribution of the boundary value (c) in our forward estimator, we get much less noise compared to the one with uniform sampling (b). (a) shows the reference estimate we get with a high sample count. (d) shows the estimate of the solution exactly on the boundary.}
\label{fig:Neumann_result}
\end{figure}

\paragraph{Boundary sampling and estimation on the boundary}
WoB for Neumann problems offers two distinct advantages over WoS, as demonstrated in Fig.~\ref{fig:Neumann_result}. First, we can begin the walk-on-boundary process by sampling a starting point on the boundary, according to the magnitude of the boundary value, to design an efficient sampling strategy when the boundary value is specified sparsely. With WoS, such a strategy is available only for Dirichlet problems~\cite{qi22bidirectional}. Second, we can exploit the BIE formulation to estimate the solution value exactly on the boundary, with a very minor modification to the underlying BIE. 
The original WoS and its basic extensions for Neumann problems that rely on epsilon shell termination criteria are incapable of estimating the solution or the normal derivative exactly on the boundary or exhibit significant bias in such cases.

\begin{figure}
    \centering
    \includegraphics[width=0.49\linewidth]{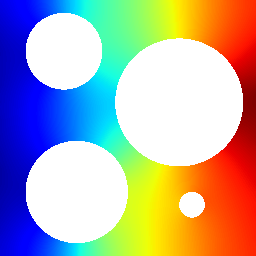}
    \includegraphics[width=0.49\linewidth]{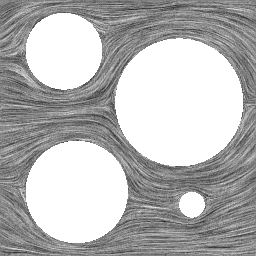}
    \caption{Potential flow reconstruction from the velocity boundary condition. We show the estimated potential field on the left and the estimated velocity field on the right. On the outer boundary, a constant inflow boundary condition $\partial u/\partial \vecn = -1$ is given on the left edge, and a constant outflow boundary condition $\partial u/\partial \vecn = 1$ is given on the right edge. On the other outer boundary edges and the inner boundaries, $\partial u/\partial \vecn = 0$ is given.}
    \label{fig:potential_flow}
\end{figure}

\paragraph{Gradient estimator and a path reuse strategy} Similarly to WoS, WoB can estimate the potential and the gradient simultaneously with almost no additional cost. Fig.~\ref{fig:potential_flow} demonstrates the use of WoB for interpolating potential flow velocities, as described by \citet{Nielsen2011}. By solving for the gradient of the scalar potential $u$, which satisfies the Laplace equation and a prescribed Neumann boundary condition, we obtain the velocity field that follows potential flow assumptions (incompressibility and irrotationality) and matches the inflow/outflow conditions at the boundaries. In this example, we use a backward estimator for the single layer formulation with RIS, in combination with a path reuse strategy analogous to the virtual point lights method~\cite{keller1997instant} in rendering; we generate sample paths from the boundaries to get samples for the unknown boundary value estimates first, and connect each of them to all the evaluation points to reuse the subpaths starting from the boundaries. This effectively increases the number of sample paths per evaluation point while introducing correlation of the estimates at different evaluation points. Though this correlation may seem undesirable, this approach guarantees that the estimated solution and gradient fields are always smooth, making it a potentially preferable alternative in some settings like fluid simulation applications.
\citet{miller2023boundary} introduced a similar boundary value caching technique for WoS in concurrent work, and their analysis largely applies to the case of WoB as well.

\begin{figure}
    \centering
    \includegraphics[width=0.32\linewidth]{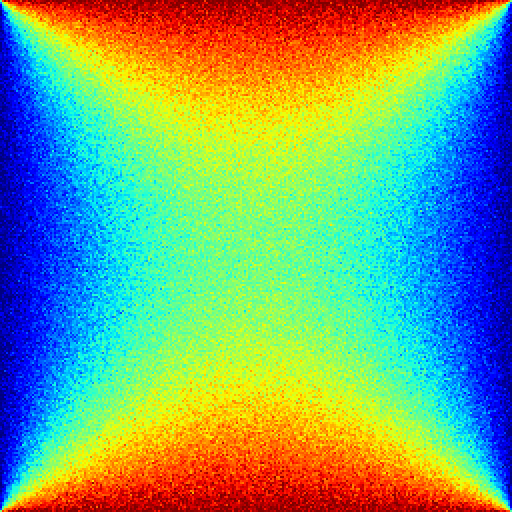}
    \includegraphics[width=0.32\linewidth]{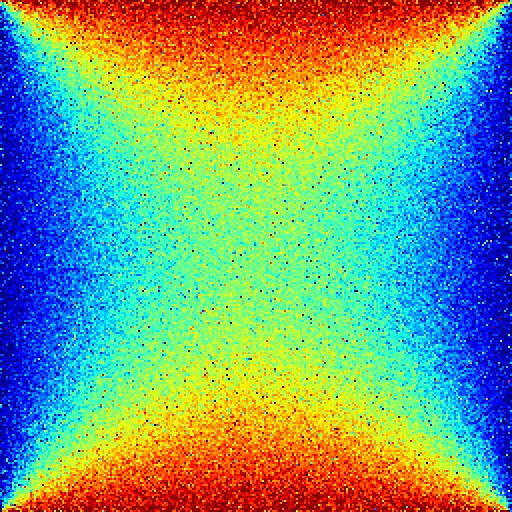}
    \includegraphics[width=0.32\linewidth]{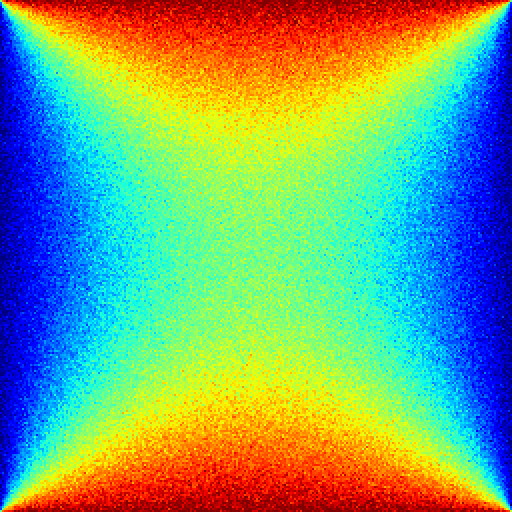}
    \includegraphics[width=0.32\linewidth]{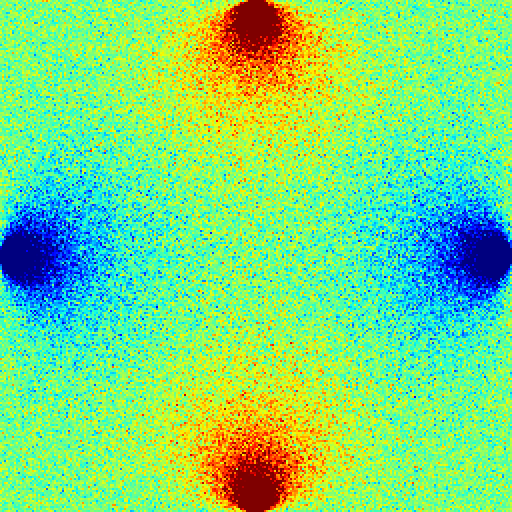}
    \includegraphics[width=0.32\linewidth]{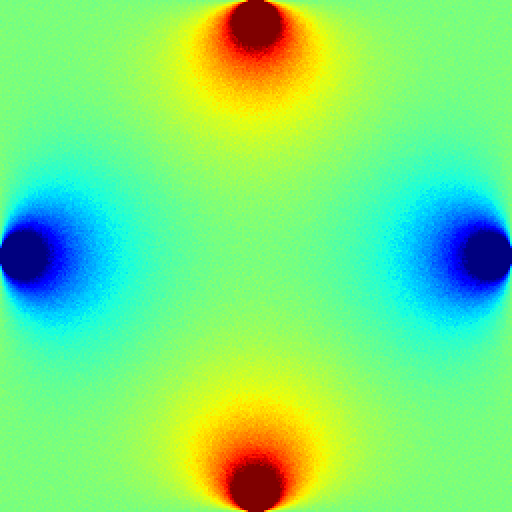}
    \includegraphics[width=0.32\linewidth]{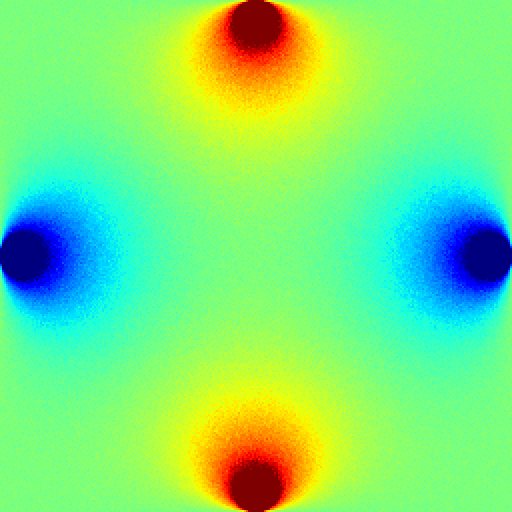}
    \caption{Equal-time comparison of bidirectional WoB estimators. Left to right: purely backward estimator, next-event estimation, and the MIS combination of the two. Top row: non-zero boundary values on each side. Bottom row: non-zero boundary values around the center of each side. Either the backward estimator or next-event estimation is more efficient than the other in each setting. WoB allows us to trivially combine the two estimators via MIS, and the combined estimator is robust across different settings (right).}
    \label{fig:bdtest}
\end{figure}

\paragraph{MIS estimator}
Due to the similarity between WoB and MC rendering, it is trivial to combine WoB estimators via MIS. 
Fig.~\ref{fig:bdtest} compares the backward estimator, the backward estimator with importance sampling of boundary values, and the MIS combination of the two estimators with the balance heuristic for a Dirichlet problem.
The two estimators correspond to unidirectional path tracing and path tracing with next-event estimation in MC ray tracing. 
When the non-zero boundary values are not localized (top row), the backward estimator performs well, although boundary sampling (which corresponds to next-event estimation) suffers from additional noise due to its explicit connection to a boundary point. 
Conversely, when the non-zero boundary values are localized (bottom row), boundary sampling becomes significantly more efficient than the backward estimator. 
This behavior is analogous to unidirectional path tracing and next-event estimation for cases where light sources are large or small. 
The MIS combination of the two estimators (right column) is robust across different settings. 
The related work for WoS~\cite{qi22bidirectional} left this MIS combination to future work, as the integration domain changes at each step in WoS (i.e., WoS solves a Volterra equation; see Appendix~\ref{sec:classes}). %
WoB allows us to incorporate MIS since it has the same mathematical and algorithmic structure as MC ray tracing.
Notably, our bidirectional estimators for WoB also do not introduce extra bias, unlike their WoS counterparts~\cite{qi22bidirectional}.

\begin{figure}
    \centering
    \includegraphics[width=0.49\linewidth]{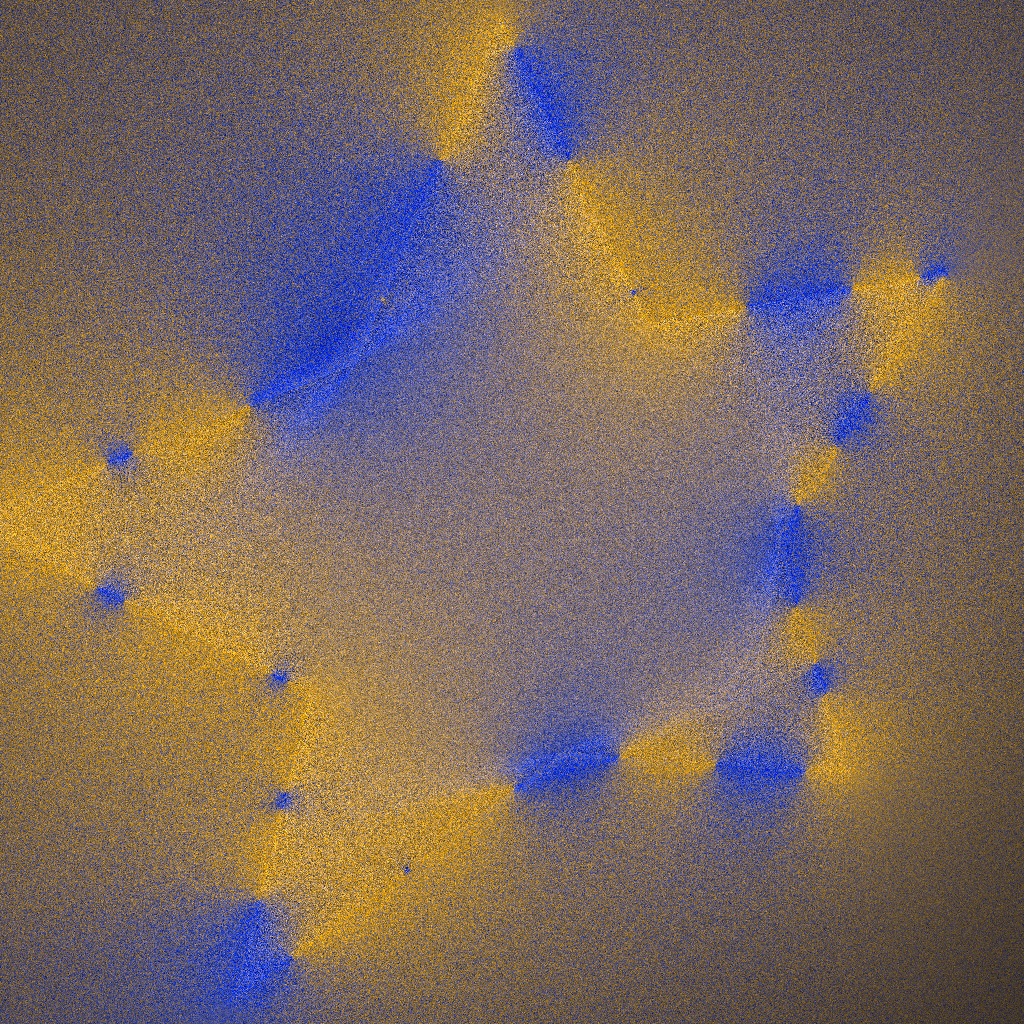}
    \includegraphics[width=0.49\linewidth]{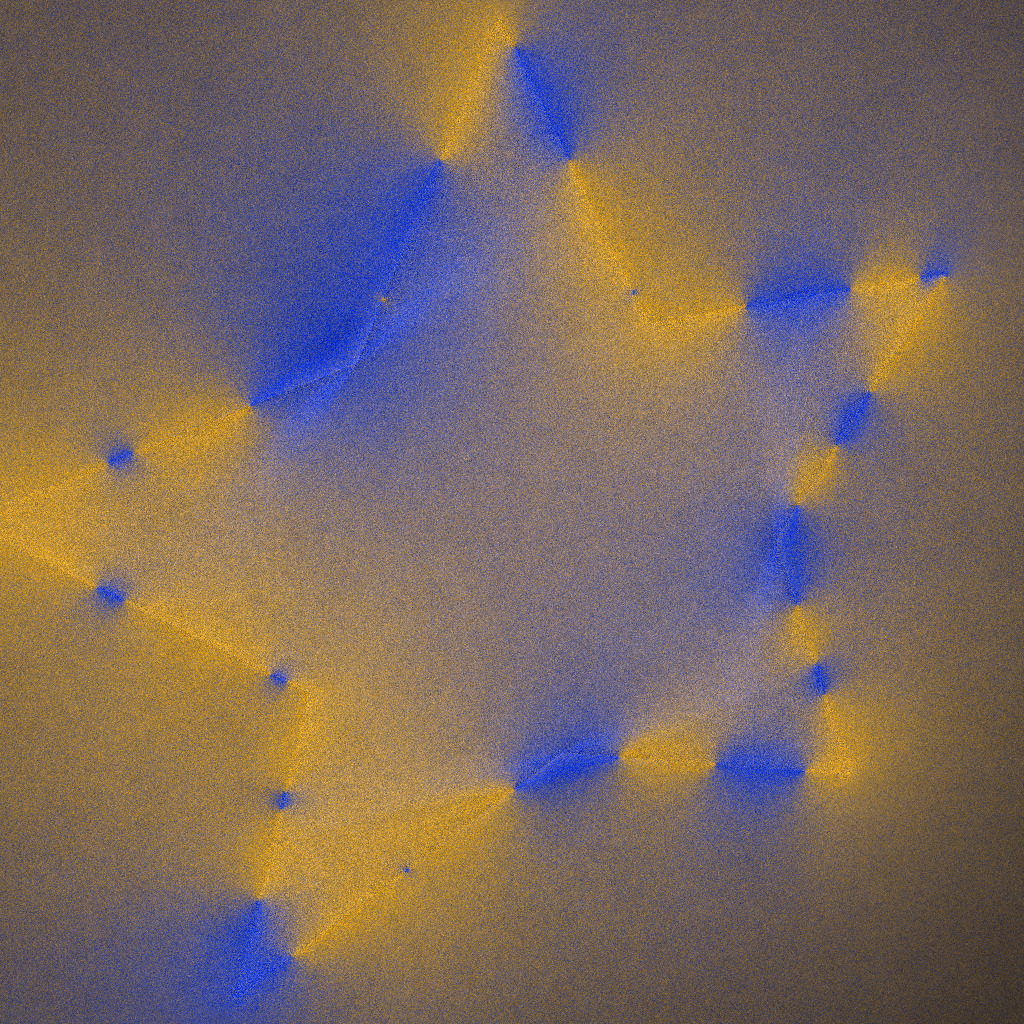}
    \caption{Equal-time comparison of Monte Carlo and Markov chain Monte Carlo (MCMC) estimators for solving interior and exterior Dirichlet problems on a plane passing through a concave 3D star shape. Left image: Monte Carlo sampling. Right image: Markov chain Monte Carlo sampling formulated as PSSMLT~\cite{Kelemen2002}. MCMC can result in lower variance than MC as in rendering.}
    \label{fig:mcmc}
\end{figure}

\paragraph{Markov chain Monte Carlo} %
We implemented primary sample space MLT~\cite{Kelemen2002} (PSSMLT) on top of WoB interior/exterior Dirichlet estimators. 
In this implementation, each sample in WoB is generated according to a Markov chain over the sampling domain including the image space. 
PSSMLT formulates this sampling domain as a unit hypercube of random numbers used to generate each sample and the location in the image space. 
In our case, the first two dimensions are used to pick a pixel (i.e., evaluation point) and the rest of the dimensions are used for generating a sample in the corresponding WoB estimator (i.e., interior or exterior, depending on the pixel), just like PSSMLT in rendering. 
We set the target distribution to be the absolute value of the sample's contribution, so samples that have higher absolute contributions are likely to be generated more often. 
Since this target distribution includes all the terms in each sample (e.g., the boundary condition and the probability of selection in all-hits), MCMC is expected to perform better than MC with a limited form of importance sampling.
Fig.~\ref{fig:mcmc} shows our preliminary examples that compare MC and MCMC in equal time. 
These examples solve interior and exterior Dirichlet problems at the same time. 
We observe that PSSMLT performs similarly in rendering and WoB in the sense that the image is less noisy than MC at the cost of correlation artifacts. 
This application of MCMC is straightforward due to the similarity between rendering and WoB.

\begin{figure}[t]
\centering
\input{Figures/WoSWoBNumerical/convex.pgf}
\hspace{-1.4cm}
\includegraphics{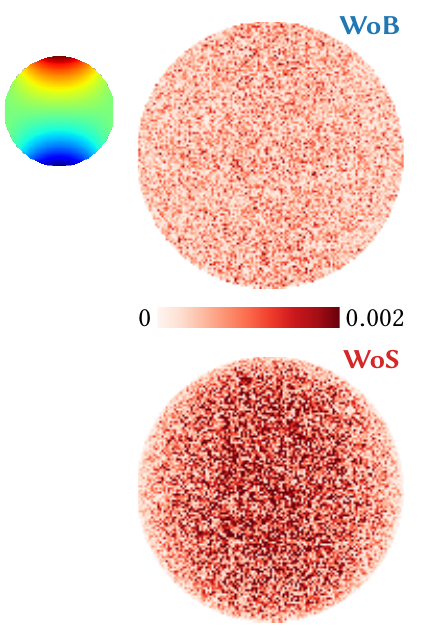}
\vspace{3mm}
\input{Figures/WoSWoBNumerical/nonconvex.pgf}
\hspace{-1.4cm}
\includegraphics{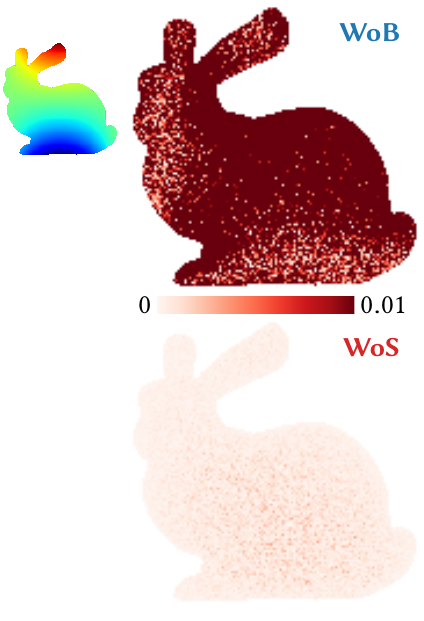}
\caption{Comparison of WoB and WoS with example interior Dirichlet problems in a convex domain (top) and a non-convex domain (bottom). The plots show how the root mean squared errors (vertical axis) decay with increasing time (horizontal axis). Each line corresponds to a specific parameter choice for WoB (blue) and WoS (red). We use path length $M=2$ to $7$ for WoB, and epsilon shell size  $10^{-2}$ to $10^{-7}$ for WoS. For each scene and for each method, we show the error of the solution with the least error after 2 hours on the right. We observe that WoS performs more efficiently than WoB for complex non-convex domains. }
\label{fig:wob_wos}
\end{figure}

\paragraph{Numerical comparisons between WoB and WoS} 
Fig.~\ref{fig:wob_wos} compares the efficiency of WoB and WoS interior Dirichlet estimators in examples of convex and non-convex domains. 
Both WoB and WoS are implemented with similarly optimized CUDA code in their basic forms with no advanced sampling techniques. 
Our experiments suggest that WoS performs more efficiently than WoB for complex non-convex domains, but comparably for simple convex domains. 
The reason is that, while WoS typically takes more computation time per sample (i.e., per individual random walk that terminates near the boundary), WoB requires more sample paths due to its higher  variance, resulting in lower overall efficiency.  We expect that this difference may become smaller with additional variance reduction techniques.
Moreover, this observation only provides general guidance on the choice between WoS and WoB. 
We do not claim that one of them is fundamentally more efficient than the other in any given problem (be it convex or non-convex) based on these numerical comparisons; the theoretical comparisons offered throughout this paper are more generally relevant (e.g., accuracy near the boundary and generality regarding supported problems).

\section{Conclusion} 
WoB is a promising alternative to WoS as an MC estimator for boundary value problems.
WoB offers a unified way to solve interior and exterior Dirichlet, Neumann, Robin, and mixed boundary problems.
WoB is the only method that can estimate solutions near or on the boundary without any spatial discretization error (as in finite element and finite difference methods) or $\epsilon$-shell error (as in WoS). 
The remarkable similarity of WoB to rendering allowed us to apply advanced MC rendering methods directly and to implement WoB atop existing ray tracing codebases.

Our paper has only scratched the surface of WoB's full potential.
Similar to WoS, WoB can be applied to other elliptic equations with known fundamental solutions by replacing the fundamental solution. 
For example, the book by \citet{SabelfeldSimonov1994} describes the application of WoB to other equations, including screened Poisson, linear elastostatics, and diffusion problems. 
We introduced WoB to graphics for the first time, but the use of BIEs has been explored in the form of conventional spatial discretization-based methods, including the boundary element method~\cite{James2006, James1999, Da2016, Hahn2015, Sugimoto:2022:BEM, Solomon2017} and the method of fundamental solutions~\cite{Martin2008}. This relationship between WoB and these methods is analogous to that between rendering techniques based on Monte Carlo methods and the radiosity method~\cite{goral1984modeling, Nishita1985Radiosity, Cohen1985Radiosity} for the rendering equation. These traditional BIE-based techniques also suggest other potential application domains of WoB. The application to problems with spatially-varying coefficients~\cite{Sawhney:2022:GFMC} is another interesting extension.

The generality of WoB in terms of the types of domains and the valid ranges of parameters (e.g., first kind equation scaling factor $k$ and Robin problem mixture weight $\overline\alpha$) require further investigation.
We focused on simply connected domains for simplicity, with the exception of the interior Neumann problem in Fig.~\ref{fig:potential_flow}. Extensions to support other multiply-connected domain problems are explained by \citet{SabelfeldSimonov1994}, but we are yet to validate their formulation and evaluate their numerical performance.

The bias of the estimator is still another topic. Instead of truncating the paths with a predefined length, we expect that other path truncation techniques, such as Russian roulette, could be applied, but with a careful investigation of how this should be done considering the modification to the Neumann series.

Lastly, we are greatly interested in improving the efficiency of the method. While WoB applies to more problems than WoS, we have not yet thoroughly evaluated the performance of WoB and the variants of WoS, where available.
However, we foresee that future applications of advanced rendering methods, such as the UPS/VCM methods~\cite{Georgiev:2012:LTS,Hachisuka:2012:PSE}, together with techniques tailored for WoB, such as the use of a more efficient modified Neumann series and more efficient handling of all-hits intersections due to contributions between mutually invisible points, can make WoB an even more attractive choice. %

\begin{acks}
This research was partially funded by NSERC Discovery Grants
(RGPIN-2021-02524 \& RGPIN-2020-03918) and a grant from Autodesk.
The second author was supported by the NSERC USRA program and the URF program at the University of Waterloo.
This research was enabled in part by support provided by SHARCNET and the Digital Research Alliance of Canada.
The 3D models in this paper are courtesy of CyberWare, Max Planck Institute, Keenan Crane, Pratt \& Whitney/Hugues Hoppe, and Stanford University.
We would like to thank the anonymous reviewers for their constructive evaluations and feedback.
\end{acks}

\bibliographystyle{ACM-Reference-Format}
\bibliography{main}

\appendix
\section{Fundamental solution}\label{app:fund_sol}%
The fundamental solution $G(\vecx, \vecy)$ for Laplace operator is the solution to the equation in an infinite domain: 
$
    \Delta G(\vecx, \vecy) + \delta(\vecx-\vecy) = 0
$. 
In 3D, we have $G(\vecx, \vecy) = \frac{1}{4\pi r}$ and its derivatives are
\begin{equation}
    \begin{split}
    &\begin{aligned}
    \frac{\partial G}{\partial \vecx_\mathsf{k}}(\vecx, \vecy) = \frac{\vecr\cdot \mathbf{e}_\mathsf{k}}{4\pi r^3}, \quad
    \frac{\partial G}{\partial \vecn_\vecy} (\vecx, \vecy) = -\frac{\vecr\cdot\vecn_\vecy}{4\pi r^3}, \quad
    \frac{\partial G}{\partial \vecn_\vecx} (\vecx, \vecy) = \frac{\vecr\cdot\vecn_\vecx}{4\pi r^3},
    \end{aligned}\\
    &\frac{\partial^2 G}{\partial \vecx_\mathsf{k} \partial\vecn_\vecy}(\vecx, \vecy) = \frac{1}{4\pi} \left[\frac{\vecn_{\vecy} \cdot \mathbf{e}_\mathsf{k}}{r^3} - 3\frac{(\vecr\cdot\vecn_\vecy)(\vecr\cdot  \mathbf{e}_\mathsf{k})}{r^5} \right].\nonumber
    \end{split}
\end{equation}
In 2D, we have $G(\vecx, \vecy) = -\frac{1}{2\pi}\log r$ and its derivatives are
\begin{equation}
    \begin{split}
    &\begin{aligned}
    \frac{\partial G}{\partial \vecx_\mathsf{k}}(\vecx, \vecy) = \frac{\vecr\cdot\mathbf{e}_\mathsf{k}}{2\pi r^2}, \quad
    \frac{\partial G}{\partial \vecn_\vecy} (\vecx, \vecy) = -\frac{\vecr\cdot\vecn_\vecy}{2\pi r^2}, \quad
    \frac{\partial G}{\partial \vecn_\vecx} (\vecx, \vecy) = \frac{\vecr\cdot\vecn_\vecx}{2\pi r^2},\\
    \end{aligned}\\
    &\frac{\partial^2 G}{\partial \vecx_\mathsf{k} \partial\vecn_\vecy}(\vecx, \vecy) = \frac{1}{2\pi} \left[\frac{\vecn_{\vecy} \cdot \mathbf{e}_\mathsf{k}}{r^2} - 2\frac{(\vecr\cdot\vecn_\vecy)(\vecr\cdot  \mathbf{e}_\mathsf{k})}{r^4} \right].\nonumber
    \end{split}
\end{equation}
where $\vecr = \vecy - \vecx$, $r= \lVert \vecr \rVert$, and $\mathbf{e}_k$ is the $k$-th basis vector.

\section{Classes of Integral Equations}\label{sec:classes}
\subsection{The rendering equation, WoS, and WoB}
Both the rendering equation and the formulation of WoS are understood to be Fredholm equations of the second kind \cite{qi22bidirectional,Pharr:2018:PBRT}. 
This type of integral equation takes the form 
$
    f(\vecx) = g(\vecx) + \int_D K(\vecx, \vecy)f(\vecy)\,d\vecy
$
where $K(\vecx, \vecy)$ is a given integral kernel, $g(\vecx)$ is a known function, $f(\vecy)$ is an unknown function we want to solve for, and $D$ is a fixed integration domain.
When the integration domain changes depending on $\vecx$, it is called a Volterra equation (of the second kind).

In rendering, for surface light transport, $\vecx$ is a tuple of a location on the surface and a direction from there (i.e., to measure radiance coming from that particular location toward the particular direction).
The function $g(\vecx)$ is the emission term, and $K(\vecx, \vecy)$ is defined as 
$
    K(\vecx, \vecy) = f_r(\vecx, \vecy) V(\vecx, \vecy) G_{eo}(\vecx, \vecy) 
$
where $f_r$ is the BSDF, $V$ is the visibility term, and $G_{eo}$ is the geometry term~\cite{Pharr:2018:PBRT}.
The integration domain is fixed as the surfaces of the scene, so it fits the definition of Fredholm equations.

In WoS, $g(\vecx)$ is defined as 
$
    g(\vecx) = \int_{B_\vecx} \overline b(\vecz) G_{B_\vecx}(\vecx, \vecz) d\vecz
$
where $\overline b(\vecx)$ is the source function, $G_{B_\vecx}(\vecx, \vecy)$ is Green's function for the largest ball $B_\vecx$ contained within the domain centered at $\vecx$. 
Note that both functions are given so $g(\vecx)$ is also still given.
The kernel $K(\vecx, \vecy)$ for WoS is defined as 
$
    K(\vecx, \vecy) = \frac{\partial G_{B_\vecx}(\vecx, \vecy)}{\partial \vecn_\vecy}
$
and the integration domain is $D=\partial B_\vecx$, the surface of the ball $B_\vecx$.
While \citet{qi22bidirectional} claimed that the formulation of WoS is a Fredholm equation of the second kind, it is a Volterra equation of the second kind since the integration domain $D=\partial B_\vecx$ changes according to $\vecx$. 
A connection between WoS and the rendering equation was imperfectly made in this sense. 
On the other hand, in WoB, the integration domain is fixed as the boundary, so it is precisely a Fredholm equation.

The rendering equation, commonly referred to as a Fredholm equation~\cite{Pharr:2018:PBRT}, appears to contradict the definition of a fixed integration domain when it is solved by MC ray tracing. 
Ray tracing from a point $\vecx$ results in a varying set of visible points $\vecy$ depending on $\vecx$, and thus it appears to be a Volterra equation.
However, the integration domain is fixed as the surfaces of the scene, so it still fits the definition of a Fredholm equation.
While this mismatch is paradoxical, we have identified that classifying the rendering equation as \emph{solely} a Fredholm equation is inaccurate.

In addition to the aforementioned area form, the rendering equation can take a solid angle form, in which the integration domain $D$ becomes the hemispherical angular domain around $\vecx$, and the kernel $K(\vecx, \vecy)$ is defined as $f_r(\vecx \rightarrow \vecy)\cos\theta$, where $\vecy$ is the first visible point from $\vecx$ along the direction towards $\vecy$~\cite{Pharr:2018:PBRT}. 
This solid-angle form is the form used in MC ray tracing, and in this context, the rendering equation is a Volterra equation of the second kind, due to the changing angular integration domain based on $\vecx$. 
However, the area form of the rendering equation, which integrates over all surface points, is a Fredholm integral equation of the second kind. 
The assertions made by \citet{qi22bidirectional} still hold if one accepts that the solid-angle form of the rendering equation is a Volterra equation, just like the formulation of WoS.

This distinction is subtle since a Volterra equation can be converted into a Fredholm equation by properly expanding the kernel $K(\vecx, \vecy)$ with zeros in a common fixed integration domain. 
This conversion is precisely what actually occurs in the area form of the rendering equation, where the visibility term $V(\vecx, \vecy)$ returns zero for points $\vecy$ that are not visible from $\vecx$. 
The area form of the rendering equation is still a Fredholm equation, but it can be transformed into a Volterra equation by redefining the integration domain to include only visible points from $\vecx$. 
The formulation of WoB cannot be reduced to a Volterra equation, as its kernel is nonzero everywhere.

\subsection{Singularity and reciprocity of the kernel}
Boundary integral equations involve singular kernels, where the kernel becomes unbounded as the distance between points $r = \lvert \vecx - \vecy \rvert$ approaches zero. 
The order of singularity in the kernel, $K$, can be classified as weakly singular, strongly singular, and hypersingular, for $K=\mathcal{O}(1/r)$, $K=\mathcal{O}(1/r^2)$, and $K=\mathcal{O}(1/r^3)$ respectively.
The orders of singularity here pertain to three-dimensional scenarios, although the underlying concepts remain unchanged in two-dimensional scenarios. 
As the order of singularity increases, robust estimation becomes challenging.
Weak and strong singularities in the kernel are prevalent in the rendering equation, and a plethora of techniques to circumvent numerical difficulties have been studied~\cite{Pharr:2018:PBRT}. 
For instance, tracing a ray, as opposed to directly sampling a point on surfaces, can avoid singularities arising from the geometry term. 
Similar methods are applied to WoB. 
However, hypersingular integrals necessitate special attention during computation and are not typically encountered in rendering. 
It is advisable to avoid high-order singularities whenever possible.

The kernel $K$ is called symmetric if $K(\vecx, \vecy) = K(\vecy, \vecx)$ and asymmetric otherwise.
In rendering, the terms reciprocal and non-reciprocal are used instead of symmetric and asymmetric, respectively;  
the kernel of the rendering equation is usually symmetric due to the physics of light, but can be asymmetric in certain cases~\cite{veach1998robust}. 
Both WoB and WoS deal with asymmetric kernels but they do not pose a problem for the solvers as long as they are treated properly, similarly to handling of asymmetric kernels in rendering.

\subsection{First-kind equations and their MC estimation}
Both Fredholm and Volterra integral equations have first and second kind forms. 
A Fredholm equation of the first kind takes the form
$
g(\vecx) = \int_D K(\vecx, \vecy)f(\vecy)\,d\vecy,
$
where both $g$ and $K$ are known and $f$ is the unknown function to be solved for. Note we used the term first kind equation abusively in the main text: strictly, we should call it a first kind equation if the equation holds for the entire integral domain, but in our mixed boundary problem estimator, we have different integral equations defined conditionally on the type of boundary at each point.
There is also a third kind of Fredholm equation, but it is not relevant to WoB in our paper.

While second-kind equations can be estimated using MC integration via Neumann series expansion, this technique is not applicable to first-kind equations. 
For instance, one might consider estimating the integral using MC integration as:
$
g(\vecx) \approx \frac1N \sum_{i=1}^N \frac{K(\vecx, \vecy_i)f(\vecy_i)}{p(\vecy_i)}.
$
This equation cannot be used, as $f$ is unknown and cannot be solved for using MC integration alone. 
We therefore focused on having second-kind equations that are compatible with MC integration, which requires us to choose a specific BIE to achieve this goal.

\end{document}